\documentclass[11pt]{article}
\usepackage{epsfig}
\usepackage{amsfonts}
\usepackage{amsmath,amssymb}
\usepackage[dvipsnames]{xcolor}
\usepackage{bbm,bm}

\usepackage{cite}
\allowdisplaybreaks[1]
\usepackage{cancel}
\usepackage{multirow}
\usepackage{tikz}

 \usepackage[
   colorlinks=true,
   linkcolor={red!50!black},
   citecolor={blue!50!black},
   filecolor=black,
   urlcolor=black,
   breaklinks=true
   ]{hyperref}
\usepackage{orcidlink}

\newcommand{\rmi}[1]{{\mbox{\scriptsize #1}}}
\newcommand{\rmii}[1]{{\mbox{\tiny\rm{#1}}}}

\newcommand{\tmu}{\bar{\mu}}

\newcommand{\g}{g}

\newcommand{\Tc}{T_{\rm c}}

\newcommand{\vw}{v_w}

\newcommand{\geff}{g_\rmi{eff}}

\newcommand{\tr}{\text{tr}}
\newcommand{\Tr}{\text{Tr}}

\newcommand{\T}{\rmii{$T$}}

\newcommand{\bG}{\widehat{G}}
\newcommand{\bA}{\hat{A}}
\newcommand{\bJ}{\hat{J}}
\newcommand{\bD}{\widehat{D}}
\newcommand{\bphi}{\hat{\phi}}

\renewcommand{\vec}[1]{{\bf #1}}
\newcommand{\nn}{\nonumber \\}
\newcommand{\gammaE}{\gamma_\rmii{E}}

\newcommand{\sumint}[1]{{\hbox{$\sum$}\!\!\!\!\!\!\!\int\,}_{\!\!\!\!\raise-0.9ex\hbox{$\scriptstyle{#1}$}}}
\newcommand{\Tint}[1]{{\hbox{$\sum$}\!\!\!\!\!\!\!\int\,}_{\!\!\!\!\raise-0.9ex\hbox{$\scriptstyle{#1}$}}}
\newcommand{\Tinti}[1]{{{\Sigma}\!\!\!\!\raise0.3ex\hbox{$\int$}_\rmii{${#1}$}}}
\newcommand{\Tintip}[1]{{{\Sigma'}\!\!\!\!\!\raise0.3ex\hbox{$\int$}_\rmii{${#1}$}}}

\makeatletter \@addtoreset{equation}{section} \makeatother
\renewcommand{\theequation}{\arabic{section}.\arabic{equation}}
\makeatletter
\renewcommand\section{\@startsection{section}{1}{\z@}%
  {-5.5ex \@plus -1ex \@minus -.2ex}%
  {2.3ex \@plus.2ex}%
  {\normalfont\large\bfseries}}
\renewcommand\subsection{\@startsection{subsection}{2}{\z@}%
  {-3.25ex\@plus -1ex \@minus -.2ex}%
  {1.5ex \@plus .2ex}%
  {\normalfont\normalsize\bfseries}}
\renewcommand\thesection{\@arabic\c@section}
\renewcommand\thesubsection{\thesection.\@arabic\c@subsection}
\renewcommand{\@seccntformat}[1]{%
  \csname the#1\endcsname.\hspace{1.0em}}
\makeatother

\begin{document}

\flushbottom

\begin{titlepage}

\begin{flushright}
  June 2026
\end{flushright}
\begin{centering}

\vfill

{\Large{\bf%
  Higher-dimensional operators and
  Polyakov loop\\
  in hot Scalar QED from the heat kernel
}}

\vspace{0.8cm}

\renewcommand{\thefootnote}{\fnsymbol{footnote}}
Siddhartha Bandyopadhyay%
\orcidlink{0009-0002-3455-643X},%
$^{\rm a,}$%
\footnotemark[1]
Joydeep Chakrabortty%
\orcidlink{0000-0001-8709-916X},%
$^{\rm a,}$%
\footnotemark[2]\\
Debmalya Dey%
\orcidlink{0009-0004-7475-9850},%
$^{\rm a,}$%
\footnotemark[3]
Philipp Schicho%
\orcidlink{0000-0001-5869-7611},%
$^{\rm b,}$%
\footnotemark[4]
Tushar%
\orcidlink{0009-0000-4057-7574},%
$^{\rm a,}$%
\footnotemark[5]

\vspace{0.8cm}
$^\rmi{a}$%
{\em
  Indian Institute of Technology Kanpur, Kalyanpur,\\
  Kanpur 208016, Uttar Pradesh, India
}
\vspace{0.3cm}

$^\rmi{b}$%
{\em
  D\'epartement de Physique Th\'eorique, Universit\'e de Gen\`eve,\\
  24 quai Ernest Ansermet, CH-1211 Gen\`eve 4, Switzerland
}
\vspace{0.3cm}

\vspace*{0.8cm}

\mbox{\bf Abstract}

\end{centering}

\vspace*{0.3cm}

\noindent
Using the finite-temperature heat kernel method,
we compute the gauge-invariant effective Lagrangian up to dimension six for massive hot scalar QED.
We propose two complementary approaches:
integrating out the non-zero Matsubara modes at finite temperature, and
deriving the finite-temperature heat kernel coefficients from their zero-temperature counterparts.
We show that in the static limit both yield
the same three-dimensional effective operators.
We also compute the gauge-invariant Coleman-Weinberg effective potential for
a constant background at finite temperature.
We further examine how the Polyakov loop modifies the matching coefficients
and assess its impact together with the
higher-dimensional operators on the thermodynamics of cosmological first-order phase transitions,
which in turn can affect
an associated gravitational-wave spectrum.

\vfill
\renewcommand{\thefootnote}{\fnsymbol{footnote}}
\footnotetext[1]{siddhartha25@iitk.ac.in}
\footnotetext[2]{joydeep@iitk.ac.in}
\footnotetext[3]{debmalyad23@iitk.ac.in}
\footnotetext[4]{philipp.schicho@unige.ch}
\footnotetext[5]{tushar25@iitk.ac.in}
\renewcommand{\thefootnote}{\arabic{footnote}}
\setcounter{footnote}{0}
\end{titlepage}

{\hypersetup{hidelinks}
\tableofcontents
}

\clearpage

\section{Introduction}
\label{sec:intro}

The early universe may have undergone cosmological first-order phase transitions (FOPTs)
that leave an observable imprint as
a stochastic gravitational-wave (GW) background~\cite{%
  Caprini:2024hue,LISACosmologyWorkingGroup:2022jok,LISA:2017pwj
  }.
The upcoming space-based interferometer LISA~\cite{Caprini:2015zlo,LISA:2017pwj}
aims to probe such backgrounds
across a broad frequency range, motivating precision calculations of
the thermodynamic parameters that characterize a FOPT,
in particular,
the transition strength $\alpha$,
the transition rate $\beta/H$, and
the bubble wall velocity $\vw$~\cite{Caprini:2019egz,Croon:2020cgk}.
Some of these parameters are most reliably computed within the framework of thermal effective
field theory (EFT), where the four-dimensional theory is reduced to a three-dimensional
EFT by integrating out the hard thermal modes at scale
$\sim \pi T$~\cite{%
  Ginsparg:1980ef,Appelquist:1981vg,Nadkarni:1988fh,Landsman:1989be,
  Kajantie:1995dw,Braaten:1995jr,Braaten:1995cm}.

The standard implementation of dimensional reduction retains only the renormalizable
operators of the three-dimensional EFT, which suffice at leading order in the
high-temperature expansion.
At subleading orders, an infinite tower of higher-dimensional operators is generated,
suppressed by powers of $1/T$, and their inclusion
becomes relevant~\cite{%
  Chala:2024xll,Bernardo:2025vkz,Chala:2025xlk,%
  Chala:2025oul,Bernardo:2026whs}
for the precision
required by next-generation GW experiments~\cite{%
  Croon:2020cgk,Gould:2021oba,Gould:2023ovu,Ekstedt:2024etx
}.
A systematic study of how such operators modify the phase-transition parameters was
recently carried out in~\cite{%
  Bernardo:2025vkz,Bernardo:2026whs},
where a dimension-six operator basis was established for
the Abelian Higgs model via diagrammatic dimensional reduction.
Reducing the operator content to a minimal, non-redundant form
requires exploiting the equations of motion (EOMs) and
field redefinitions to
eliminate physically equivalent operators.

An alternative and algorithmically powerful route to the one-loop thermal effective
action is provided by the heat kernel expansion
in Schwinger time~\cite{Schwinger:1951nm,DeWitt:1975ys}
which was carried out in the context of
QCD~\cite{Dyakonov:1984,Chapman:1994vk,Megias:2003ui}, and
recently generalized
to generic thermal effective actions~\cite{Chakrabortty:2024wto,Balui:2025yvd}.
This approach is manifestly gauge-covariant and generates the full tower of
higher-dimensional operators in a closed form.
In the present work, we apply this method to hot scalar QED,
the massive vector Abelian gauge theory coupled to a massive complex scalar,
and systematically extract the dimension-six operator basis of its finite-temperature EFT.
Field redefinitions need to be employed
to reduce the resulting operators to a minimal,
gauge-invariant form
similar to the one obtained in~\cite{Chakrabortty:2026swu}
for the Standard Model effective theory (SMEFT).

We compare the operator basis and matching coefficients obtained via the heat kernel
against the diagrammatic results of~\cite{Bernardo:2025vkz} and
the software package {\tt DRalgo}~\cite{Ekstedt:2022bff,Bernardo:2026nyq},
finding agreement in the static limit and
after locally expanding non-local operators generated by the heat kernel.
As a further extension, we examine how the Polyakov loop modifies the matching
coefficients, an effect not studied earlier,
and identify it as a new non-perturbative input parameter.
Its dependence on the gauge charge of the particles makes
it potentially significant in analyses of
cosmological phase transitions and
associated gravitational-wave spectra.

The paper is organized as follows.
Sec.~\ref{sec:eft} reviews the thermal EFT framework and describes
the heat kernel approach to dimensional reduction.
In sec.~\ref{sec:HKEA},
we first introduce the Abelian Higgs model and fix our conventions,
then we introduce
two approaches to obtain the heat kernel coefficients at finite temperature.
The heat kernel construction of the dimension-six operator basis is carried out
in sec.~\ref{sec:AH:heatKernel}. We present the one-loop effective Lagrangian and the Coleman-Weinberg (CW) potential at finite temperature.
Effects
of the Polyakov loop on phase-transition thermodynamics
are analyzed in sec.~\ref{sec:PolyakovLoop:effects}. In sec.~\ref{sec:dr}, we discuss the compatibility of the heat kernel method with the diagrammatic one. 
Conclusions and an outlook are given in sec.~\ref{sec:outlook}.
Master integrals and heat kernel coefficients are collected in the
appendices~\ref{sec:masterIntegrals} and~\ref{sec:heatKernel coefficients}.

\section{Effective field theory at finite temperature}
\label{sec:eft}

At zero temperature and in Euclidean spacetime,
the one-loop effective action is defined on
the $D$-dimensional Euclidean manifold
\begin{align}
  \label{eq:zeroT:manifold}
  \mathcal{M}_{\T = 0} = \mathbb{R}^{1} \times \mathbb{R}^{d}
  \,,
\end{align}
with $D = d+1$ and $d = 3$,
where
the momenta take continuous values $P_\mu \in \mathbb{R}^{D}$ and
we define $X \equiv (x_0, x^i)$ as the Euclidean spacetime coordinate.
Finite temperature $\beta \equiv 1/T$ is introduced by compactifying
the Euclidean time direction on a circle $S^1_\beta$ of circumference $\beta$
and imposing (anti-)periodic boundary conditions on
the generic fields $\Phi$,
\begin{equation}
  \label{eq:PBC}
  \Phi(x_0 + \beta,\, \mathbf{x}) = (-1)^\sigma\,\Phi(x_0,\, \mathbf{x})\,,
\end{equation}
where
$\sigma = 0$ for bosons (periodic) and
$\sigma = 1$ for fermions (anti-periodic).
The topology of the Euclidean manifold thereby changes to
\begin{align}
  \label{eq:finiteT:manifold}
  \mathcal{M}_{\T} = S^1_\beta \times \mathbb{R}^{d}
   \,,
\end{align}
while the spatial directions remain non-compact.

This compactification explicitly breaks the $O(D)$ Euclidean
rotational symmetry of zero-temperature theory down to $O(d)$,
singling out the temporal direction as physically distinct from the
$d$ spatial ones.
The $D$-dimensional covariant derivative $D_\mu$,
with $\mu \in \{0, i\}$, accordingly decomposes as
\begin{equation}
  \label{eq:Dmudecomp}
  D_\mu \;\longrightarrow\; (D_0,\, D_i)\,,
\end{equation}
where the spatial components $D_i$ ($i = 1, \ldots, d$) retain the residual
$O(d)$ symmetry and play the role of the covariant derivative
of the dimensionally reduced theory, while the temporal component $D_0$
is a singlet under $O(d)$.

The compactification also has a direct consequence for the momentum spectrum. 
While in $\mathbb{R}^{D}$,
the temporal component $p_0$ is integrated over
continuously, in $\mathbb{R}^d \times S^1_\beta$
the boundary condition given in eq.~\eqref{eq:PBC}
discretizes it to the Matsubara frequencies~\cite{Matsubara:1955ws},
\begin{align}
  \label{eq:Matsubara}
  p_0 &\longrightarrow
    \omega_n = \bigl(2n + \sigma\bigr)\pi T
  \,,&
  n &\in \mathbb{Z}
  \,,
\end{align}
where the integration $\int \frac{{\rm d}p_0}{2\pi}$ is replaced by the
discrete sum $T\sum_{n}$.

\section{Effective action using the heat kernel}
\label{sec:HKEA}

Before specializing on finite temperature in sec.~\ref{sec:heatKernel:finiteT},
we first review the heat kernel method for
computing the one-loop effective action at zero temperature.

The one-loop effective action is determined by
the functional determinant of the fluctuation operator obtained from
the quadratic expansion of the action around a classical background.
We define $\Delta$ as the strong elliptic operator, in the Euclidean space, given by
the second functional derivative of the Euclidean action $(\mathcal{S})$ with respect to
the quantum fluctuations $\Phi$ 
\begin{equation}
  \label{eq:elliptic_op}
 \Delta_{ij}(X,Y) \equiv 
  \frac{\delta^2\mathcal{S}}{\delta\Phi_i(X)\,\delta\Phi_j(Y)}
  \Bigg|_{\substack{%
      \Phi_i(X)=0\\
      \Phi_i(Y)=0}}
  =   \Big[\bigl(D^2 + M^2\bigr)_{ij} + U_{ij}^{ }\Big] (X,Y)
  \,,
\end{equation}
where $M^2$ is the mass matrix and $U$ encodes 
all interaction terms.
The covariant derivative $D_\mu$ acts on the 
fields in the appropriate representation of the gauge group.

We now focus on the Euclidean formulation of
the heat kernel $K(t,X,Y,\Delta)$~\cite{%
  Seeley:1967ea,Belkov:1995gjw,Vassilevich:2003xt,Avramidi:2001ns,
  Kontsevich:1994xe,Banerjee:2023iiv,Banerjee:2023xak,Chakrabortty:2023yke}
.
This
is defined as the solution to
the heat equation in
the Euclidean manifold $\mathcal{M}_{\T=0}$ eq.~\eqref{eq:zeroT:manifold},
\begin{equation}
  \label{eq:heat_eq}
  \bigl(\partial_t + \Delta_X\bigr)\,K(t,X,Y,\Delta) = 0
  \,,
\end{equation}
where
$t$ is the proper time parameter
in the Schwinger proper-time representation~\cite{Fock:1937dy,Schwinger:1951nm,DeWitt:1975ys}.
The initial condition is
$K(0,X,Y,\Delta) = \delta^{(D)}(X-Y)$,
where $X \equiv (x_0, x^i)$ and $Y \equiv (y_0, y^i)$
are full $D$-dimensional Euclidean spacetime coordinates.

The trace of the heat kernel $\tr\,K(t,X,X,\Delta)$ encodes
the local spectral information of the operator $\Delta$.
Here,
$\Tr\, \mathcal{O}$ denotes the full functional trace over both spacetime and internal indices, while
$\tr\, \mathcal{O}$ denotes the trace over internal indices only.
Using the Schwinger proper-time representation, 
the functional trace of the logarithm
can be written as~\cite{Banerjee:2023iiv, Banerjee:2023xak,Chakrabortty:2023yke}
\begin{equation}
  \Tr\log\,\Delta 
  = -\int_0^\infty \frac{{\rm d}t}{t}\,\Tr\,e^{-t\Delta}
  = -\int_0^\infty \frac{{\rm d}t}{t}\int_X\tr\,K(t,X,X,\Delta)\,,
\end{equation}
where
$\int_X \equiv \int {\rm d}^{D}X$ is the spacetime integral
with $D$ being the spacetime dimension,
so that the one-loop effective Lagrangian takes the form
\begin{equation}
  \label{eq:HKLeff}
  \mathcal{L}_{\rmi{eff}} 
  = c_s\int_0^\infty \frac{{\rm d}t}{t}\;\tr\,K(t,X,X,\Delta)
  \,,
\end{equation}
where
$c_s = 1/2$ for real and
$c_s = +1$ for complex scalars,
encoding the degeneracy of the one-loop functional determinant.
The ultraviolet divergences of the theory are captured by the
small-$t$ behavior of the heat kernel~\cite{Vassilevich:2003xt}.
Here, we primarily focus on computing the finite contributions
to the local effective operators.

The heat kernel admits a momentum-space representation \cite{Banerjee:2023xak}
\begin{align}
  \label{eq:THK}
  \tr\, K(t,X,X,\Delta) 
  &= \tr\int_P\,
     \bigl\langle X
     \big|\,e^{-M^2 t}\,\mathcal{T}\exp\biggl[
       -\int_0^t\!\Bigl(D^2 + e^{M^2 t'}\,U\,e^{-M^2 t'}\Bigr)\,{\rm d}t'
     \biggr]
     \big|P
     \bigr\rangle
     \bigl\langle P
     \big|X
     \bigr\rangle
     \nn
  &= \tr\int_P
     e^{-M^2 t}\,e^{P^2 t}\;
     \mathcal{T}\exp\biggl[
       -\int_0^t\!\Bigl(
            D^2
         + 2i P\cdot D
         + e^{M^2 t'}\,U\,e^{-M^2 t'}
       \Bigr)\,{\rm d}t'
     \biggr]
    \,,
\end{align}
where
$\int_P \equiv \int\frac{{\rm d}^{D}P}{(2\pi)^{D}}$,
$\mathcal{T}$ denotes path ordering in the Schwinger parameter $t$,
and the trace runs over internal indices.
Here
$P_\mu$ is the loop momentum and
$D_\mu$ carries the background-field information.
For non-degenerate masses,
$e^{M^2 t'}U\,e^{-M^2 t'}$ produces
exact exponential entries $e^{\pm\Delta_{12}^2 t'}$ on
each off-diagonal insertion
(cf.\ eq.~\eqref{eq:AH:heatKernel:At})
yielding a result that is exact in the mass splittings
$\Delta_{12}^2$.
Rescaling $P \to P/\sqrt{t}$,
eq.~\eqref{eq:THK} takes the following form:
\begin{equation}
  \tr\,K(t,X,X,\Delta) 
  = \tr\int_P \frac{e^{-M^2 t}\,e^{P^2}}{t^{\frac{D}{2}}}\,
    \;
    \mathcal{T}\exp\biggl[
      -\int_0^t\!\Bigl(
          D^2
        + \frac{2i P\cdot D}{\sqrt{t}} 
        + e^{M^2 t'}\,U\,e^{-M^2 t'}
      \Bigr)\,{\rm d}t'
    \biggr]
    \,,
\end{equation}
where $P^2 = \eta_{\mu\nu}P^\mu P^\nu =
-(p_1^2 + \dots + p_4^2)$ in Euclidean signature.

To compute the heat coefficients for
non-degenerate masses,
we define
\begin{equation}
  \mathcal{F}(t,\mathcal{A}) 
  = \mathcal{T}\exp\Bigl(-\int_0^t\mathcal{A}(t')\,{\rm d}t'\Bigr)
  = 1 + \sum_{n=1}^{\infty}(-1)^n\,f_n(t,\mathcal{A})
  \,,
\end{equation}
where the $f_n$ are nested Volterra integrals,
\begin{align}
  \label{eq:hk_fn}
  f_n(t,\mathcal{A}) 
  = \int_0^t {\rm d}s_1
    \int_0^{s_1}{\rm d}s_2\cdots
    \int_0^{s_{n-1}}{\rm d}s_n\;
    \mathcal{A}(s_1)\,\mathcal{A}(s_2)\cdots\mathcal{A}(s_n)
  \,.
\end{align}
For a system of two non-degenerate fields
$(\Phi_1,\Phi_2)$ 
with mass splitting
$\Delta^2_{12} = M_1^2 - M_2^2 = -\Delta^2_{21}$,
the integrand matrix $\mathcal{A}(t')$ takes the block form, e.g.,
for two non-degenerate fields,
\begin{equation}
  \label{eq:AH:heatKernel:At:non-degenerate}
  \mathcal{A}(t') =
  \begin{pmatrix}
      D^2
    + \frac{2i P\cdot D}{\sqrt{t}}
    + U_{11} 
    &
    U_{12}\,e^{\Delta_{12}^2 t'} \\[2mm]
    U_{21}\,e^{\Delta_{21}^2 t'} 
    &
      D^2
    + \frac{2i P\cdot D}{\sqrt{t}}
    + U_{22}
  \end{pmatrix}
  \,,
\end{equation}
where the entries $U_{ij}$ with $i,j\in \{1,2\}$
contain the interactions among the fields.
Substituting into eq.~\eqref{eq:HKLeff},
the one-loop effective 
Lagrangian in four-dimensional Euclidean space is expressed compactly as
\begin{equation}
  \label{eq:Leff:final}
  \mathcal{L}_{\rmi{eff}} 
  = c_s\;\tr\int_0^\infty
    \frac{{\rm d}t}{t}
    \frac{e^{-M^2 t}}{t^{D/2}}
    \int_P
    e^{P^2}\,
    \biggl[
        1
      + \sum_{n=1}^\infty(-1)^n f_n(t,\mathcal{A})
      \biggr]
    \,.
\end{equation}

where $f_n$ are the non-degenerate analogues of the zero temperature degenerate heat kernel coefficients.

\subsection{Scalar QED: Effective Lagrangian}
\label{sec:model}

We consider the Abelian Higgs model with
a ${\rm U}(1)$ complex scalar field $\phi$ and
gauge field $A_\mu$ at finite temperature $T$.
The corresponding Lagrangian in Minkowski spacetime reads
\begin{equation}
  \label{eq:AH:lagrangian}
  \mathcal{L} =
      |D_\mu\phi|^2
    - M_1^2\phi^\dagger \phi
    - \frac{\lambda}{6} (\phi^\dagger \phi)^2
    - \frac{1}{4} G_{\mu\nu}G^{\mu\nu}
    + \frac{1}{2} M_2^2 A_\mu^2
    \,,
\end{equation}
where
$\phi(x) \equiv \frac{1}{\sqrt{2}} (\phi_1 + i \phi_2)$,
$D_\mu = \partial_\mu + iA_\mu$
is the covariant derivative,
$G_{\mu \nu}$ is
the Abelian gauge field tensor, and
$\g$ is the $U(1)$ gauge coupling.\footnote{Here, we absorb $\g$ within $A_\mu$.}
To implement the heat kernel method, we will henceforth work in Euclidean space. 
The fields are expanded around their classical backgrounds as
$\phi_a = \bphi_a + h_a(x)$ and
$A_\mu = \bA_\mu + \eta_\mu(x)$,
where $h_a(x)$ and $\eta_\mu(x)$ are the scalar and gauge-field fluctuations,
respectively.
Then the background gauge field $\bA_\mu$  appears in the background covariant derivative
$\bD_\mu = \partial_\mu + \bA_\mu$. Here, we work with background Fermi gauge: $- \frac{1}{2\xi}(\bD^\mu \eta_\mu)^2$.
Collecting the fluctuation fields into
a field multiplet $\Phi = (h_a,\, \eta_\mu)^T$,
the elliptic operator, see eq.~\eqref{eq:elliptic_op}, reads
\begin{align}
   \Delta_{\alpha \beta} &=
\begin{pmatrix}
  \bigl[ \bD^2 + M_1^2 \bigr] \delta_{ab}
  &  0\\[4pt]
  0
  & - \!\bigl[ \bD^2 + M_2^2\bigr] \eta_{\mu\nu}
    + \!\bigl(1 - \tfrac{1}{\xi} \bigr) \bD_\mu \bD_\nu
  \end{pmatrix}
  + U
  \,,
\end{align}
where
$(\alpha,\beta)=\{ (a,b),(\mu,\nu)\}$.
It has been noted in~\cite{Balui:2025kat, Balui:2025yvd, Balui:2026ghs}
that the heat kernel method relying on the background field method
along with the background Fermi gauge provides a gauge invariant and
gauge parameter independent effective Lagrangian and therefore the potential at
zero and finite temperatures.
Thus, for simplicity and without loss of generality,
we set $\xi=1$.
The matrix $U$ contains the non-derivative and single-derivative operators,
coming from the potential.
Next to $U$,
we also define a matrix $U'$ that contains only non-derivative operators,
{\em viz.}
\begin{align}
U &=
  \begin{pmatrix}
    U'_{ab}  & \g \epsilon_{ab'} \bD_\nu \bphi_{b'} \\[2pt]
    - \g \epsilon_{a'b} \bD_\mu \bphi_{a'}   & U'_{\mu\nu}
  \end{pmatrix}
\,, &
U' &=
  \begin{pmatrix}
  U'_{ab} & 0 \\
  0 & U'_{\mu\nu}
  \end{pmatrix}
  \,,
\end{align}
with
$U'_{ab} =
    \frac{\lambda}{6} \bphi^2 \delta_{ab}^{ }
  + \frac{\lambda}{3} \bphi_a \bphi_b$, 
$U'_{\mu\nu} =
  - \g^2 \bphi^2 \eta_{\mu \nu}^{ }=
  \mbox{diag} ( 0, - \g^2 \bphi^2 \eta_{ij}^{ })$.%
\footnote{%
  The massive gauge field has
  three physical polarizations, which
  is encoded in the calculation even when
  the gauge field has a field-dependent mass.
}

At zero temperature,
the one-loop effective Lagrangian admits the standard heat kernel expansion
\begin{align}\label{zerotempL}
  \mathcal{L}_{\rmi{eff}} =
  & c_s \int_0^\infty \frac{{\rm d}t}{t}\, \frac{1}{(4\pi t)^\frac{d}{2}}
  \bigg\{
  \nn[1mm] &
    + e^{-M_1^2 t} \Bigl[
      \bigl(f^\rmii{$S$}_0 + f^\rmii{$SG$}_0\bigr)
    - t\bigl(f^\rmii{$S$}_1 + f^\rmii{$SG$}_1\bigr)
    + \tfrac{t^2}{2}\bigl(f^\rmii{$S$}_2 + f^\rmii{$SG$}_2\bigr)
    - \tfrac{t^3}{3!}\bigl(f^\rmii{$S$}_3 + f^\rmii{$SG$}_3\bigr)
    \Bigr]
  \nn &
  + e^{-M_2^2 t} \Bigl[
      \bigl(f^\rmii{$G$}_0 + f^\rmii{$GS$}_0\bigr)
    - t\bigl(f^\rmii{$G$}_1 + f^\rmii{$GS$}_1\bigr)
    + \tfrac{t^2}{2}\bigl(f^\rmii{$G$}_2 + f^\rmii{$GS$}_2\bigr)
    - \tfrac{t^3}{3!}\bigl(f^\rmii{$G$}_3 + f^\rmii{$GS$}_3\bigr)
    \Bigr]
  \bigg\}
  \,,
\end{align}
where $c_s$ is the overall prefactor from the one-loop functional-determinant formula,
encoding the boson/fermion sign and any degeneracy factor.
The explicit forms of
$f^{\rmii{$S$}}_n$,
$f^{\rmii{$SG$}}_n$,
$f^{\rmii{$G$}}_n$, and
$f^{\rmii{$GS$}}_n$
are listed in appendix~\ref{sec:heatKernel coefficients}.
In this notation, we have separated
the $f_n$ as
$f^{\rmii{$S$}}_n$ and
$f^{\rmii{$SG$}}_n$ which divides the whole expression into
the degenerate result and non-degenerate mixing effects.
Similarly for the gauge sector,
we have $f^\rmii{$G$}_n$ and $f^\rmii{$GS$}_n$.

It is important to note that although we are working
in an all-negative Euclidean convention,
all the indices from here onward are contracted using the positive Euclidean metric.

\subsection{%
  Heat kernel coefficients:
  Matching from zero to finite temperature
  }
\label{sec:AH:heatKernel:zeroT}

The heat kernel coefficients at finite temperature $(\tilde{f})$ can be obtained by
matching the zero-temperature heat kernel expansion coefficients $(f)$ to the finite-temperature one.
This strategy is detailed in~\cite{Chakrabortty:2024wto}.

The matching procedure for the scalar and vector sectors takes the form
\begin{align}\label{hkczeros}
    \sum^\infty_{k=0}\Big\{
          f^{\rmii{$S$}}_k
        + f^{\rmii{$SG$}}_k
      \Big\} \frac{(-t)^k}{k!}
      \Bigg|_{\hspace{0.05cm}(U\to U-Q^2)} &=
      e^{Q^2t}\sum^\infty_{m=0}\Big\{
          \tilde{f}^{\rmii{$S$}}_m
        + \tilde{f}^{\rmii{$SG$}}_m
      \Big\} \frac{(-t)^m}{m!}
    \nn &=
    \sum^\infty_{n,m=0} \frac{(Q^2 t)^n}{n!}\Big\{
        \tilde{f}^{\rmii{$S$}}_m
      + \tilde{f}^{\rmii{$SG$}}_m\Big
      \} \frac{(-t)^m}{m!}
    \nn & =
    \sum^\infty_{n,m=0} (-1)^m\frac{(Q^2 )^n}{n!\,m!}\Big\{
        \tilde{f}^{\rmii{$S$}}_m
      + \tilde{f}^{\rmii{$SG$}}_m
      \Big\} t^{n+m}
    \,,
\end{align}
\begin{align}\label{hkczerog}
       \sum^\infty_{k=0}\Big\{
        f^{\rmii{$G$}}_k
      + f^{\rmii{$GS$}}_k\Big\} \frac{(-t)^k}{k!} \Bigg|_{\hspace{0.05cm}(U\to U-Q^2)}
      &= \sum^\infty_{n,m=0} (-1)^m\frac{(Q^2 )^n}{n!\,m!}\Big\{ \tilde{f}^{\rmii{$G$}}_m + \tilde{f}^{\rmii{$GS$}}_m\Big\} t^{n+m}
    \,,
\end{align}
where the matching condition $U\rightarrow U-Q^2$ reads as 
$U_{ab} \to U_{ab}-\delta_{ab} Q^2$ and $U_{\mu\nu} \to U_{\mu\nu}+\eta_{\mu\nu}Q^2$. Since the matching relation is linear in the heat kernel coefficients,
the pure-sector contributions
($f^\rmii{$S$}_k$, $f^\rmii{$G$}_k$)
and the mixing contributions
($f^\rmii{$SG$}_k$, $f^\rmii{$GS$}_k$)
can be matched independently, preserving the separation between
degenerate and non-degenerate effects.
The corresponding heat kernel coefficients at finite temperature are
listed in the appendix~\ref{app:Z-THKC}.

\subsection{%
  Heat kernel coefficients:
  Directly integrating out at finite temperature
  }
\label{sec:heatKernel:finiteT}

In the finite-temperature manifold $\mathcal{M}_{\T}$ as given in
eq.~\eqref{eq:finiteT:manifold},
the heat kernel
is defined as~\cite{Megias:2003ui, Moral-Gamez:2011wcb, Chakrabortty:2024wto}
\begin{align}
\tr \, K(t; X, X, \Delta) &=
    \tr \, \frac{1}{\beta} \sum_{p_0} \biggl(
        \frac{e^{-M^2 t}}{t^{d/2}} \int_{\vec{p}} e^{-|\vec{p}|^2} \, \Big[ 1 + \sum_n (-1)^n f_n(t, \mathcal{A}) \Big]
        \biggr)
  \,,
\end{align}
where $\int_\vec{p} \equiv \int\frac{{\rm d}^d p}{(2\pi)^d}$
is the $d$-dimensional spatial momentum integral,
$f_n(t, \mathcal{A})$ are the nested Volterra integrals of eq.~\eqref{eq:hk_fn},
and $p_0$ runs over the Matsubara frequencies eq.~\eqref{eq:Matsubara}.
Compared to the zero-temperature eq.~\eqref{eq:Leff:final},
the full $D$-dimensional loop integral factorizes into a spatial Gaussian integral
and a discrete Matsubara sum.
The latter integrates out
the heavy modes and generates the thermal Wilson coefficients.

The auxiliary operator matrix $\mathcal{A}(t')$ entering the Volterra integral
takes the following form for the Abelian Higgs model,
\begin{align}
  \label{eq:AH:heatKernel:At}
\mathcal{A}(t') &=
  \begin{pmatrix}
  \left[ -Q^2 + \bD^2 + \frac{2 i p \cdot \bD}{\sqrt{t}} \right] \delta_{ab}  & 
  \bigl(\g \, \varepsilon_{ab'} Q\, \bphi_{b'} \bigr) e^{\Delta_{12}^2 t'} & 
  \bigl(\g \, \varepsilon_{ab'} \bD_i \bphi_{b'}  \bigr) e^{\Delta_{12}^2 t'}
  \\[2mm]
  -\bigl( \g \, \varepsilon_{a'b} Q\,\bphi_{a'} \bigr) e^{-\Delta_{12}^2 t'} & 
  -\left[ -Q^2 + \bD^2 + \frac{2 i p \cdot \bD}{\sqrt{t}} \right] \eta_{00} & 0
  \\[2mm]
  -\bigl( \g\, \varepsilon_{a'b} \bD_j \bphi_{a'}  \bigr) e^{-\Delta_{12}^2 t'} & 0 &
  -\left[ -Q^2 + \bD^2 + \frac{2 i p \cdot \bD}{\sqrt{t}} \right] \eta_{ij}
  \end{pmatrix}
  \nn[3mm] &
  + U'
  \,,
\end{align}
where
now momenta are purely spatial,
$Q = \bD_0 + i p_0$,
$\Delta^2_{12} = M_1^2 - M_2^2$, the metric is
$\eta_{\mu\nu} = \mbox{diag}(\eta_{00}, \eta_{ij}) = -\delta_{\mu\nu}$ and $\varepsilon$ is the anti-symmetric Levi-Civita symbol.
Henceforth we define
$\bD^2 \equiv -\sum^d_{i=1} \bD_i \bD_i$,
with the number of spatial dimensions $d=3$.

After performing the Volterra integrals,
and recovering the $f_n$ integrals,
we follow the strategy of~\cite{Chakrabortty:2024wto}
and extend the matching of the heat kernel coefficients
at the operator level.
The block-diagonal prefactor
$\mathrm{diag}(e^{-M_1^2 t}\delta_{ab},\, e^{-M_2^2 t}\delta_{00},\, e^{-M_2^2 t}\delta_{ij})$
is common to both sides and cancels,
leaving the sector-wise matching condition
\begin{align}
  \label{eq:hk:matching}
  \sum_k \frac{(-t)^k}{k!}\, B^k_X
  &= \sum_{k_1,\, k_2}
     \frac{(-1)^{k_1}\,(Q^2)^{k_2}}{k_1!\; k_2!}\,
     \widetilde{B}^{k_1}_X\; t^{k_1+k_2}
  \,, \qquad X \in \{ab,\, 00,\, ij\}\,,
\end{align}
where
\begin{align}
  B^N_X &= \delta_X^{ } + \Bigl(e^{s_X \Delta_{12}^2 t} - 1\Bigr)
  \sum_{n=2}^{4} (-1)^n \widetilde{C}^{[n,N]}_X
  \,,
\end{align}
with
$s_{ab} = +1$,
$s_{00} = s_{ij} = -1$, and $N=2,\dots,6$.
The $\widetilde{C}^{[n,N]}_X$ functions
are listed in appendix~\ref{app:D-THKC}.

\section{Effective action at finite temperature: scalar QED}
\label{sec:AH:heatKernel}

Now we apply the heat kernel construction to the Abelian Higgs model
and extract the dimension-six operator basis of the high-temperature EFT.
The matching condition in
eqs.~\eqref{hkczeros}, \eqref{hkczerog},
and~\eqref{eq:hk:matching} implies that
\begin{align}\label{eq:B-tilde:f-tilde}
  \tilde{B}^n_s &= \tilde{f}^\rmii{$S$}_n + \tilde{f}^\rmii{$SG$}_n
  \,, \nn
  \tilde{B}^n_g &= \tilde{f}^\rmii{$G$}_n + \tilde{f}^\rmii{$GS$}_n
  \,, \quad n = 0, 1, 2, 3 \,,
\end{align}
where the explicit heat kernel coefficients $\tilde{f}^\rmii{$X$}_n$ are given
in the appendix~\ref{sec:heatKernel coefficients}.


The effective Lagrangian can be written as
\begin{align}
  \label{eq:AH:heatKernel:Leff}
  \mathcal{L}_{\rmi{eff}}^{\rmi{1-loop}} &=
  \frac{1}{2} \bigg\{
    \tilde{B}_{s,\mathbf{0}}^{0}\,I^{M_1}_{\Omega}\bigl(0;0\bigr)
  + \tilde{B}_{g,\mathbf{0}}^{0}\,I^{M_2}_{\Omega}\bigl(0;0\bigr)
  - \tilde{B}_{s,\mathbf{0}}^{1}\,I^{M_1}_{\Omega}\bigl(0;1\bigr)
  - \tilde{B}_{g,\mathbf{0}}^{1}\,I^{M_2}_{\Omega}\bigl(0;1\bigr)
  \nn[2mm] &
  \hphantom{{}\frac{1}{2}\bigg[}
  + \frac{1}{2!} \Big[
      \tilde{B}_{s,\mathbf{0}}^{2}\,I^{M_1}_{\Omega}\bigl(0;2\bigr)
    + \tilde{B}_{s,\mathbf{1}}^{2}\,I^{M_1}_{\Omega}\bigl(1;\tfrac{3}{2}\bigr)
    + \tilde{B}_{g,\mathbf{0}}^{2}\,I^{M_2}_{\Omega}\bigl(0;2\bigr)
    + \tilde{B}_{g,\mathbf{1}}^{2}\,I^{M_2}_{\Omega}\bigl(1;\tfrac{3}{2}\bigr)
    \Big]
  \nn[2mm] &
  \hphantom{{}\frac{1}{2}\bigg[}
  - \frac{1}{3!} \Big[
      \tilde{B}_{s,\mathbf{0}}^{3}\,I^{M_1}_{\Omega}\bigl(0;3\bigr)
    + \tilde{B}_{s,\mathbf{1}}^{3}\,I^{M_1}_{\Omega}\bigl(1;\tfrac{5}{2}\bigr)
    + \tilde{B}_{s,\mathbf{2}}^{3}\,I^{M_1}_{\Omega}\bigl(2;2\bigr)
    + \tilde{B}_{s,\mathbf{3}}^{3}\,I^{M_1}_{\Omega}\bigl(3;\tfrac{3}{2}\bigr)
  \nn[1mm] &
  \hphantom{\frac{1}{2}\bigg[\frac{1}{3!}\Big[}
    + \tilde{B}_{g,\mathbf{0}}^{3}\,I^{M_2}_{\Omega}\bigl(0;3\bigr)
    + \tilde{B}_{g,\mathbf{1}}^{3}\,I^{M_2}_{\Omega}\bigl(1;\tfrac{5}{2}\bigr)
    + \tilde{B}_{g,\mathbf{2}}^{3}\,I^{M_2}_{\Omega}\bigl(2;2\bigr)
    + \tilde{B}_{g,\mathbf{3}}^{3}\,I^{M_2}_{\Omega}\bigl(3;\tfrac{3}{2}\bigr)
  \Big]
\bigg\}
\,.
\end{align}
The coefficients $\tilde{B}^{k}_{s/g,\mathbf{n}}$ are given in eq.~\eqref{eq:B-tilde:f-tilde} and
the sum of the Matsubara frequencies is defined in appendix~\eqref{eq:heatKernel:IOmega}.

The temporal gauge field $\bA_0$ satisfies
the following equation $\frac{{{\rm d}}\bA_0}{{\rm d}\tau}=0$,
which implies that $\bA_0$ is a function of coordinates of
$\mathbb{R}^3$.
The usual choice of gauge at finite temperature is
$\bA_0=0$, which is inconsistent as it does not guarantee
the removal of non-redundant states and
also is not compatible with the periodic boundary condition of
$\bA_i$ in the thermal partition function~\cite{Weiss:1980rj}.
Here, we work with the gauge $\bA_0={\rm const}$.
At finite temperature,
we define the Polyakov loop~\cite{%
  Polyakov:1975rs, Weiss:1980rj,Megias:2003ui,Moral-Gamez:2011wcb}
as
\begin{equation}
\label{eq:polyakov}
   \Omega (\vec{x}) =
  \text{Tr}\;\;  \mathbb{P}\, \Big[\exp \Big(-\int_{0}^{\beta}\! {\rm d}\tau \;\bA_0(\tau,\vec{x}) \Big) \Big]
		\stackrel{\bA_0(\tau,\vec{x})=\bA_0}{=}
		e^{-\beta \bA_0}
	\,,
\end{equation}
where $\mathbb{P}$ denotes the path-ordering and the trace is over the gauge indices.

For a field $\bphi_a$ charged under
$\mathrm{U}(1)$, the temporal covariant derivative is
$\bD_0 \bphi_a = (\partial_0 + \bA_0) \bphi_a$, where $\bA_0$ is constant in time.
In our work, the contribution of the Polyakov loop is captured in
the thermal Wilson coefficients,
and it constitutes
master sums
$I_\Omega$
of eq.~\eqref{eq:heatKernel:IOmega}
and
$S_\Omega$
of eq.~\eqref{eq:heatKernel:SOmega:compact}.
In this way,
the integer Matsubara modes are changed by a real (non-integer) number
depending on the gauge charge of the infrared fields,
\begin{align}
  \label{eq:polyakov:shift}
   p_0 &= 2\pi n T \to 2\pi (n + \tilde{n}) T
   \,,&
  \tilde{n} &= \frac{i}{2\pi} \langle\ln \Omega \rangle \in \mathbb{R}
  \,,
\end{align}
where the average is taken over the gauge states.
In the static limit
$(Q\bphi_a)=\bA_0 \bphi_a $,
the one-loop effective Lagrangian thus takes the following form
\begin{align}
\label{eq:Lef-nondeg}
\mathcal{L}_{\rmi{eff}}^{{\rmi{1-loop}}} &=
  - \frac{M_1^4}{32\pi^2} \Bigl( \ln\frac{M_1^2}{\tmu^2} - \frac{3}{2} \Bigr) + S_\Omega^{M_1}[0;0]
  - \frac{3M_2^4}{64\pi^2} \Bigl( \ln\frac{M_2^2}{\tmu^2} - \frac{5}{6} \Bigr) + \frac{3}{2}S_\Omega^{M_2}[0;0]
  \nn[1mm] &
  - \biggl[
      \frac{\lambda}{3}\bphi^2
    + {\frac{\g^2}{2 \Delta_{12}^2} \Bigl( \hat{A}_0^2\hat{\phi}^2}+ (\bD_i \bphi_a)(\bD_i \bphi_a)\Bigr)
  \biggr]
  \biggl[
      \frac{M_1^2}{(4\pi)^2} \Bigl(\ln\frac{M_1^2}{\tmu^2} - 1\Bigr)
    + S_\Omega^{M_1}[0;1]
  \biggr]
  \nn[1mm] &
  - \biggl[
      \frac{3}{2}\g^2\bphi^2
    {- \frac{\g^2}{ 2\Delta_{12}^2} \Bigl( \hat{A}_0^2\hat{\phi}^2}+(\bD_i \bphi_a)(\bD_i \bphi_a)\Bigr) 
  \biggr]
  \biggl[
      \frac{M_2^2}{(4\pi)^2} \Bigl(\ln\frac{M_2^2}{\tmu^2} - 1\Bigr)
    + S_\Omega^{M_2}[0;1]
  \biggr]
  \nn[1mm] &
  + \frac{1}{4} \biggl[
        \frac{5\lambda^2}{18}\bphi^4
      + \frac{2}{3}\bigl(2E_i^2 + E_{ii0}\bigr)
      + \frac{1}{3}\bG_{ij} \bG_{ij}
    \biggr]
    \biggl[ - \frac{1}{(4\pi)^2}\ln\frac{M_1^2}{\tmu^2} + S_\Omega^{M_1}[0;2] \biggr]
  \nn[1mm] &
  - \frac{1}{3}E_{ii}\,I_\Omega^{M_1}\bigl(1;\tfrac{3}{2}\bigr) + \frac{3\g^4}{4}\bphi^4
    \left[ -\frac{1}{(4\pi)^2}\ln\frac{M_2^2}{\tmu^2} + S_\Omega^{M_2}[0;2] \right]
  \nn[1mm] &
  + \frac{\g^2\lambda}{2(\Delta_{12}^2)^2}
    \biggl[
        \frac{1}{2}(\bD_i \bphi_a)(\bD_i \bphi_a)\bphi^2
      - \frac{1}{12} (\bD_i \bphi^2)^2
    +{\frac{1}{6}\hat{A}_0^2\hat{\phi}^4}\biggr]
  \nn[1mm] &
  \hphantom{{} \frac{\g^2\lambda}{(\Delta_{12}^2)^2}}
    \times
    \biggl[
      - \frac{\Delta_{12}^2}{(4\pi)^2}
      + \frac{M_2^2}{(4\pi)^2}\ln\frac{M_1^2}{M_2^2}
      + S_\Omega^{M_1}[0;1]
      - S_\Omega^{M_2}[0;1]
      + \Delta_{12}^2 S_\Omega^{M_1}[0;2]
    \biggr]
  \nn[1mm] &
  - \frac{\g^4}{2(\Delta_{12}^2)^2}(\bD_i\bphi_a)(\bD_i\bphi_a)\bphi^2
  \nn[1mm] &
  \hphantom{{} \frac{\g^2\lambda}{(\Delta_{12}^2)^2}}
     \times
    \biggl[
        - \frac{\Delta_{12}^2}{(4\pi)^2}
        + \frac{M_1^2}{(4\pi)^2}\ln\frac{M_1^2}{M_2^2}
        + S_\Omega^{M_1}[0;1]
        - S_\Omega^{M_2}[0;1]
        + \Delta_{12}^2 S_\Omega^{M_2}[0;2]
    \biggr]
  \nn[1mm] &
  - \biggl[
      \frac{1}{12}\Bigl(-
          \frac{2\lambda}{3} (\bD_i \bphi^2) E_{i0}
        + 2E_{i00}E_i
        + E_{i0}^2
        + \frac{\lambda}{3}\bphi^2(\bG_{ij})^2
        - \frac{1}{5}\bJ_i^2
      \Bigr)
      \nn[1mm] &
      \hphantom{{}-\Bigl[}
      + \frac{\lambda^2}{108}(\bD_i \bphi_a)(\bD_i \bphi_a)\bphi^2
      + \frac{\lambda^2}{108}(\bD_i \bphi^2)^2 + \frac{7\lambda^3}{648}\bphi^6
    \biggr]
    \biggl[ \frac{1}{(4\pi)^2 M_1^2} + S_\Omega^{M_1}[0;3] \biggr]
  \nn[1mm] &
  - \biggl[
      \frac{\g^6}{4}\bphi^6
      + \frac{\g^4}{8}(\bD_i \bphi^2)^2
    \biggr]
    \biggl[ \frac{1}{(4\pi)^2 M_2^2} + S_\Omega^{M_2}[0;3] \biggr]
  \nn[1mm] &
  - \frac{1}{12}\Bigl[\frac{4 \lambda}{3}(\bD_i \bphi^2) E_i - 2 E_{i0}E_i - 2E_iE_{i0}\Bigr]
    I^{M_1}_{\Omega}\bigl[1;\tfrac{5}{2}\bigr]
  \nn[1mm] &
  - \frac{1}{12}\Bigl[8E_i^2 + 2E_{ii0}\Bigr]
    I^{M_1}_{\Omega}\bigl(2;2\bigr)
  + \frac{1}{3}E_{ii}\,I^{M_1}_{\Omega}\bigl[3;\tfrac{3}{2}\bigr]
  \,,
\end{align}   
where
the tree-level Lagrangian is given in eq.~\eqref{eq:AH:lagrangian}, and
$\tmu$ is the $\overline{\text{MS}}$ renormalization scale.
Here, the electric component of the field tensor is
$E_i=\bG_{0i}=-[\bD_i,\bD_0]$,
$E_{ii}=[\bD_i,E_i]$,
$E_{i0}=[Q,E_i]$,
$E_{i00}=[Q,[Q,E_i]]$ and the magnetic component is
$\bG_{ij}=[\bD_i,\bD_j]$.
In general, $\bA_0$ is a function of the coordinates of
$\mathbb{R}^3$.
However one can certainly choose a gauge $\bA_0=\text{const.}$ that implies the vanishing of electric fields as
$E_i \sim \bD_i(\bA_0)=0$.
For the Abelian gauge symmetry,
the term
${\bG_{ij} \bG_{jk} \bG_{ki}}$ vanishes identically due
to the Bianchi identity, which justifies the absence of this operator in eq.~\eqref{eq:Lef-nondeg}.

In the degenerate mass limit of $M_1 \rightarrow M_2 \equiv M $,
the one-loop effective Lagrangian simplifies to
\begin{align}
\mathcal{L}_{\rmi{eff}}^{\rmi{1-loop}} &=
  - \frac{M^4}{32 \pi^2} \Bigl( \ln\frac{M^2}{\tmu^2} - \frac{3}{2} \Bigr)
  - \frac{3 M^4}{64 \pi^2} \Bigl( \ln\frac{M^2}{\tmu^2} - \frac{5}{6} \Bigr) + \frac{5}{2}  S_\Omega^{M}[0;0]
  \nn[1mm] &
    - \Bigl(
        \frac{\lambda}{3}
      + \frac{3\g^2}{2}
      \Bigr)\bphi^2 \left[
        \frac{M^2}{(4\pi)^2} \Bigl(\ln\frac{M^2}{\tmu^2} - 1 \Bigr)
      +  S_\Omega^{M}[0;1]
  \right]
  \nn[1mm] &
  + \biggl[
        {\frac{\g^2}{2}\Big( \hat{A}_0^2\hat{\phi}^2} +(\bD_i \bphi_a)(\bD_i \bphi_a) \Big)
      + \frac{1}{12} \bG_{ij} \bG_{ij}
      + \frac{1}{6}\bigl( 2 E_i^2 + E_{ii0} \bigr)
      + \Bigl(
          \frac{5\lambda^2}{72}
        + \frac{3 \g^4}{4} 
      \Bigr)\bphi^4
    \biggr]
    \nn[1mm] &
    \hphantom{{}-\Bigl[}
    \times
    \left[ - \frac{1}{(4\pi)^2} \ln\frac{M^2}{\tmu^2} + S_\Omega^{M}[0;2] \right]
    - \frac{1}{3} E_{ii}\, I_\Omega^{M}\bigl(1;\tfrac{3}{2}\bigr)
  \nn[1mm] &
  - \biggl[
      \frac{1}{12}\Bigl( -
          \frac{2\lambda}{3} (\bD_i \bphi^2) E_{i0}
        + 2 E_{i00} E_i
        + E_{i0}^2
        + \frac{\lambda}{3} \bphi^2 (\bG_{ij})^2
        - \frac{1}{5} \bJ_i^2
      \Bigr)
      \nn &
      \hphantom{-\Bigl[}
      + \frac{\g^2\lambda}{24}\hat{A}_0 ^2 \hat{\phi}^4
      + \Bigl(
            \frac{\g^4}{4}
          + \frac{\g^2\lambda}{8}
          + \frac{\lambda^2}{108}
        \Bigr) (\bD_i \bphi_a) (\bD_i \bphi_a) \, \bphi^2
      \nn[1mm] &
      \hphantom{-\Bigl[}
      + \Bigl(
            \frac{\g^4}{8}
          - \frac{\g^2 \lambda}{48}
          + \frac{\lambda^2}{108}
        \Bigr) (\bD_i \bphi^2)^2
      + \Bigl(
        \frac{7\lambda^3}{648} 
      + \frac{\g^6}{4}
      \Bigr)\bphi^6
    \biggr]
    \left[ \frac{1}{(4\pi)^2 M^2} + S_\Omega^{M}[0;3] \right]
    \nn &
  - \frac{1}{12} \Bigl[\frac{4 \lambda}{3}(\bD_i \bphi^2) E_i - 2 E_{i0} E_i - 2 E_i E_{i0} \Bigr]
    I^{M}_{\Omega}\bigl(1;\tfrac{5}{2}\bigr)
  \nn[1mm] &
   - \frac{1}{12} \Bigl[ 8 E_i^2 + 2 E_{ii0} \Bigr]
    I^{M}_{\Omega}\bigl(2;2\bigr)
  + \frac{1}{3} E_{ii}\, I^{M}_{\Omega}\bigl(3;\tfrac{3}{2}\bigr)
  \,.
\end{align}
The degenerate mass limit is smooth, and the effective Lagrangian is free of any singularity.
The effective Lagrangian in the degenerate mass limit can also be obtained by directly applying
the heat kernel construction for degenerate masses, as given in eq.~\eqref{eq:Lef-nondeg}.
The consistency between the two approaches provides a non-trivial check of our results.

\subsection{Coleman-Weinberg effective potential}

The heat kernel method can also be used to compute
the Coleman-Weinberg effective potential~\cite{Coleman:1973jx}.
We assume that both the scalar and gauge fields are massless in the tree-level Lagrangian.
They only receive masses after expansion around a constant background,
so their masses depend on the background fields.
Following this proposal,
we identify the $M^2$ and $U$ matrices as
\begin{align}
M^2 &=
  \begin{pmatrix}
		M_1^2\delta_{ab} & 0 \\
		0 & - M^2_2 \eta_{\mu \nu} 
	\end{pmatrix}
  \,,&
    U &= \begin{pmatrix}
		\frac{\lambda}{3} \bphi_a \bphi_b - \frac{\lambda}{3} \bphi^2 \delta_{ab} & \g \epsilon_{ab^\prime} \bD_\nu \bphi_{b^\prime}\\
		-\g \epsilon_{a{^\prime}b} \bD_\mu \bphi_{a^\prime} & 0
	\end{pmatrix}
  \,,
\end{align}
where
$M_1^2 = \frac{\lambda}{2} \bphi^2$ and
$M_2^2 = \g^2 \bphi^2$ are
the field-dependent masses of the scalar and gauge fields, respectively.
By applying
the definition of the matrix $\mathcal{A}(t)$ in eq.~\eqref{eq:AH:heatKernel:At:non-degenerate},
we can express $\mathcal{A}(t)$ as
\begin{align}
    \mathcal{A}(t') &= \begin{pmatrix}
        \Bigl[\bD^2 + \frac{2ip\cdot\bD}{\sqrt{t}} - \frac{\lambda}{3} \bphi^2\Bigr]\delta_{ab}
      + \frac{\lambda}{3} \bphi_a \bphi_b
      &
        \g \epsilon_{ab^\prime} \bD_\nu \bphi_{b^\prime} e^{\Delta_{12}^2 t'}
      \\
      - \g \epsilon_{a{^\prime}b} \bD_\mu \bphi_{a^\prime} e^{-\Delta_{12}^2 t'}
      &
      - \Bigl[\bD^2 + \frac{2ip\cdot\bD}{\sqrt{t}} \Bigr] \eta_{\mu \nu} 
   \end{pmatrix}
   \,.
\end{align}

Inserting the matrices $M^2$, $U$, $\mathcal{A}$ in
eq.~\eqref{eq:HKLeff} and
restricting to the local effective Lagrangian,
we obtain
the one-loop Coleman-Weinberg (CW) effective potential~\cite{Coleman:1973jx}
\begin{align}
	V_{\rmii{CW}} ( \bphi) &=
  \frac{1}{64 \pi^2} \tr\bigg\{
      M_1^4 \Big( \ln\frac{M_1^2}{\tmu^2} -\frac{3}{2} \Big)
    + M_2^4 \Big( \ln\frac{M_2^2}{\tmu^2} -\frac{5}{6} \Big)
    + U_{11}^2 \ln\frac{M_1^2}{\tmu^2}
    + U_{22}^2 \ln\frac{M_2^2}{\tmu^2}
  \nn[1mm] &
  \hphantom{\frac{1}{64 \pi^2} \tr\bigg\{}
    + 2M_1^2 U_{11} \Big( \ln\frac{M_1^2}{\tmu^2} -1 \Big)
	  + 2M_2^2 U_{22} \Big( \ln\frac{M_2^2}{\tmu^2} -1 \Big)
    \bigg\}
  \nn[1mm] &=
  \frac{1}{4!}\frac{1}{8 \pi^2} \biggl[
      \frac{5}{6} \lambda^2 \bphi^4 \Big(\ln\frac{\lambda \bphi^2}{2\tmu^2} - \frac{3}{2}\Big) 
    + 9 \g^4 \bphi^4 \Big(\ln\frac{\g^2 \bphi^2}{\tmu^2} -\frac{5}{6}\Big)
    \biggr].
\end{align}
Here, we assume the background fields to be constant and thus
they do not have any anomalous dimensions.
This leads to the absence of the term
$V_\rmii{CW} \supset \g^2\lambda\,\bphi^4$ in the CW potential~\cite{Balui:2026ghs},
as this contribution emerges through
the kinetic term $(\bD_i \bphi_a)(\bD_i \bphi_a)$ of
the background field.

A complete description of the phase transition also requires the effective action.
In this context,
higher-order corrections to the kinetic term of
the scalar field background appear in eq.~\eqref{eq:Lef-nondeg} and
will modify both the bounce action and the nucleation rate.
The impact of such corrections on the bounce action
has been studied in~\cite{Chala:2024xll}.

\subsection{%
  Finite-temperature potential and
  Polyakov loop effects}
\label{sec:PolyakovLoop:effects}

At finite temperature,
the integrand matrix 
of eq.~\eqref{eq:AH:heatKernel:At} that enters
the Volterra integration in eq.~\eqref{eq:hk_fn}
reflects the absence of Lorentz symmetry and
the temporal direction of the gauge field becomes
explicitly present.
Following the approach of
secs.~\ref{sec:AH:heatKernel:zeroT}
and~\ref{sec:heatKernel:finiteT},
we construct the finite-temperature
potential
\begin{align}
  \label{eq:Veff:CW:finiteT}
  V_{\rmii{CW}}^{\beta} ( \bphi )&=
	    V_{\rmii{CW}} (\bphi)
  - \text{tr}\bigg\{
        \frac{1}{2}S_{\Omega}^{M_1} [ 0;0]
      + \frac{1}{2}S_{\Omega}^{M_2} [ 0;0]
      + U_{11} \frac{1}{2}S_{\Omega}^{M_1} [ 0;1]
      + U_{22} \frac{1}{2}S_{\Omega}^{M_2} [ 0;1]
    \nn[1mm] &
    \hphantom{{}V_{\rmii{CW}} (\bphi)+ \text{tr}\bigg\{}
      + \frac{1}{4} U_{11}^2 S_{\Omega}^{M_1} [ 0;2]
      + \frac{1}{4} U_{22}^2 S_{\Omega}^{M_2} [ 0;2]
    \bigg\}
    \nn[2mm] &=
	    V_{\rmii{CW}} (\bphi)
    - S_{\Omega}^{\frac{\lambda}{2} \bphi^2} [ 0;0]-\frac{3}{2}S_{\Omega}^{ \, \g^2 \bphi^2} [ 0;0]
    + \frac{\lambda}{3!}\bphi^2 S_{\Omega}^{\frac{\lambda}{2} \bphi^2} [ 0;1]
    - \frac{\lambda^2}{36}\bphi^4 S_{\Omega}^{ \,\frac{\lambda}{2} \bphi^2} [ 0;2]
    \,.
\end{align}

To quantify the phase transition thermodynamics,
we need to evaluate the free energy difference between the symmetric and broken phases.
To this end,
we define
\begin{align}
  \label{eq:vacuum:subtraction}
  \Delta V^{\beta}(\bphi)
    &= V^{\beta}(\bphi) - V^{\beta}(0)
  \,,&
  \Delta V^{\beta}(0) &= 0
  \,,
\end{align}
where
$V^{\beta}_{ }(\bphi) = V^{\beta}_{\rmi{tree}}(\bphi) + V_{\rmii{CW}}^{\beta}(\bphi)$.
Since we are interested
in the qualitative features
of higher-dimensional operators and the Polyakov loop effects,
we first fix a benchmark (BM) point in the $(\lambda,\g)$ plane,
\begin{align}
  \label{eq:BM1}
  \lambda &= 0.05
  \,,&
  \g &= 0.8
  \,,
  \tag{BM1}
\end{align}
with the couplings run
from a reference scale $\tmu_{\rmi{ref}}$
to the thermal scale $\tmu = \pi \hat{T}\,\tmu_{\rmi{ref}}$,
where $\hat{T} = T/\tmu_{\rmi{ref}}$,
via the corresponding $\beta$-functions.
See e.g.~\cite{Hirvonen:2021zej,Bernardo:2025vkz}
for the explicit expressions of the $\beta$-functions in
the Abelian Higgs model.

\begin{figure}[t]
  \centering
  \includegraphics[width=0.5\textwidth]{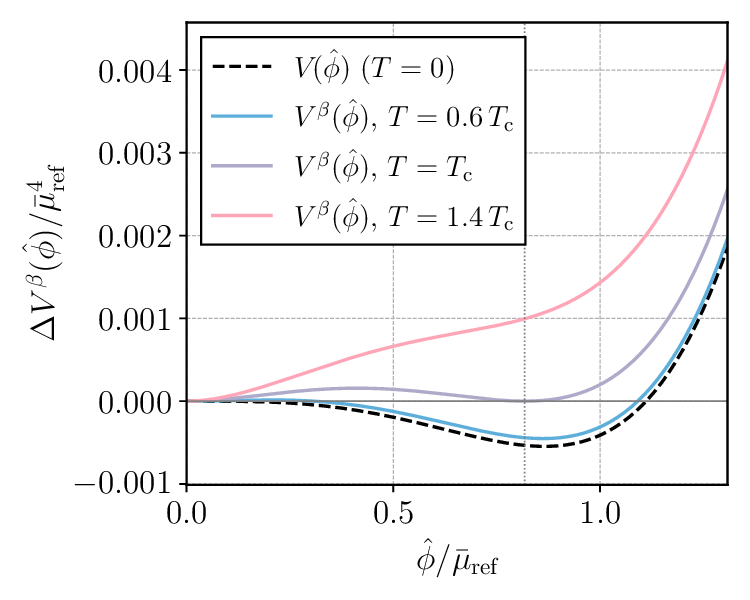}%
  \includegraphics[width=0.5\textwidth]{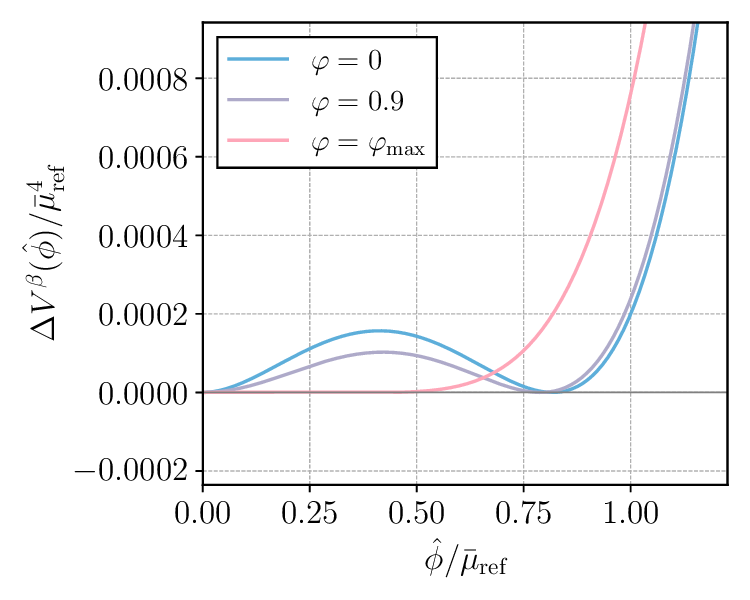}
  \caption{%
    The finite-temperature free energy difference
    $\Delta V^{\beta}_{ }(\bphi)$ of
    eq.~\eqref{eq:vacuum:subtraction}
    at different temperatures (left) and
    different values of the Polyakov phase $\varphi$ (right).
    Left:
    At the critical temperature $T=\Tc$ (solid), the symmetric and broken
    minima are degenerate and separated by a barrier, signalling a
    first-order transition;
    for $T>\Tc$ only the symmetric minimum survives.
    The dashed curve is the zero-temperature Coleman-Weinberg potential
    $V_{\rmii{CW}}(\bphi)$.
    Right:
    For larger values of the Polyakov phase $\varphi$,
    the thermal corrections are suppressed and
    the critical temperature $\Tc$ is higher.
    At $\varphi_{\rmi{max}} \approx 1.32$,
    the transition becomes second order.
  }
  \label{fig:Veff:finiteT}
\end{figure}
Figure~\ref{fig:Veff:finiteT} displays
the zero-temperature
Coleman-Weinberg potential together with its finite-temperature counterpart
$V^{\beta}(\bphi)$.
Thermal corrections, encoded in the master sums $S_\Omega$,
generate a barrier between the symmetric and broken phases.
The degeneracy of the two minima defines the critical
temperature $\Tc \approx 0.21\,\tmu_{\rmi{ref}}$ for the benchmark point~\eqref{eq:BM1}.

The thermal corrections
in eq.~\eqref{eq:Veff:CW:finiteT}
go beyond a naive Matsubara sum at the trivial Polyakov loop holonomy.
In the heat kernel construction, the temporal background
$\bA_0$ enters the master sums $S_\Omega$ and $I_\Omega$ through the
Polyakov loop $\Omega$ in eq.~\eqref{eq:polyakov}.
The latter shifts the integer Matsubara frequencies
$\omega_n = 2\pi n T$ by a
real amount set by the gauge charge of the infrared field.
The Polyakov loop enters the Matsubara sum through a gauge-space
averaging in eq.~\eqref{eq:polyakov:shift},
which is why it appears as a number dressing the master
sums rather than as an operator-valued tower of $\bA_0^n$ insertions
as in the diagrammatic dimensional reduction.
See sec.~\ref{sec:dr:differences} for a detailed discussion of the differences.
The heat kernel approach keeps the full holonomy dependence intact, which
affects thermal screening,
and consequently also the phase-transition thermodynamics.

To be more quantitative,
we introduce the Polyakov phase
$\varphi \equiv 2\pi\tilde{n} \bmod 2\pi \in [0,2\pi)$,
with $\tilde{n} = \frac{i}{2\pi} \langle\ln \Omega \rangle$,
and vary it at the benchmark point~\eqref{eq:BM1},
tracking its imprint on
the effective potential in fig.~\ref{fig:Veff:finiteT}.
By increasing $\varphi$ from the trivial-holonomy value $\varphi=0$,
the thermal corrections are suppressed,
the critical temperature $\Tc$ increases, and
the transition becomes weaker until at
$\varphi_\rmi{max} \approx 1.32$,
the transition turns second order.

Additionally, we investigate the impact of the Polyakov loop on
the critical temperature $\Tc$ and
the transition strength
$\alpha = \Delta \theta/\rho_{\rmi{rad}}$.
Here,
$\theta \equiv T^\mu_{\ \mu} = e - 3p$
is the trace anomaly,%
\footnote{%
  For gravitational-wave applications, $\alpha$ is more
  faithfully related to the pseudotrace
  $\bar\theta \equiv e - p/c_{s,\text{bro}}^2$.
  For a general broken-phase sound speed $c_{s,\text{bro}}$,
  the pseudotrace correctly predicts
  the energy converted into bulk fluid motion, and hence the resulting
  gravitational-wave spectrum~\cite{Giese:2020rtr,Giese:2020znk}.
}
with
$e$ being the energy density,
$p$ the pressure,
$\rho_{\rmi{rad}} = \pi^2 \geff T^4/30$ the radiation energy density, and
$\geff$ the effective number of relativistic degrees of freedom.
\begin{figure}[t]
  \centering
  \includegraphics[width=0.5\textwidth]{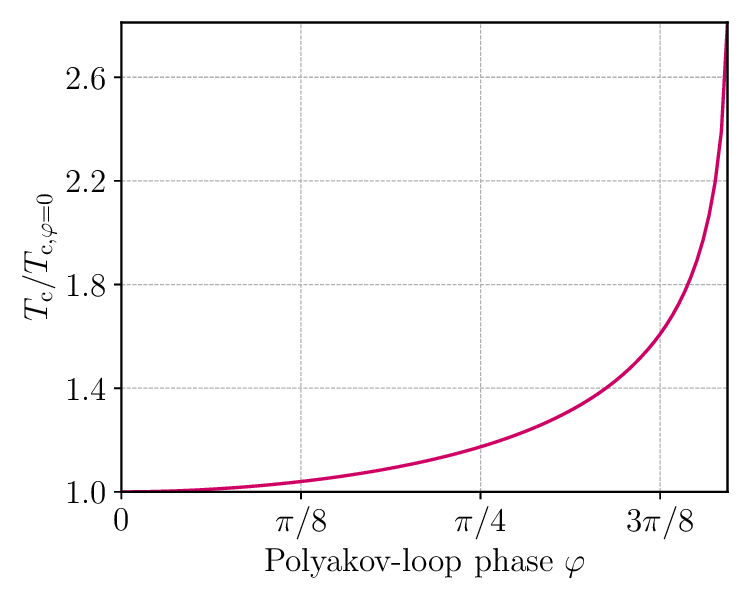}%
  \includegraphics[width=0.5\textwidth]{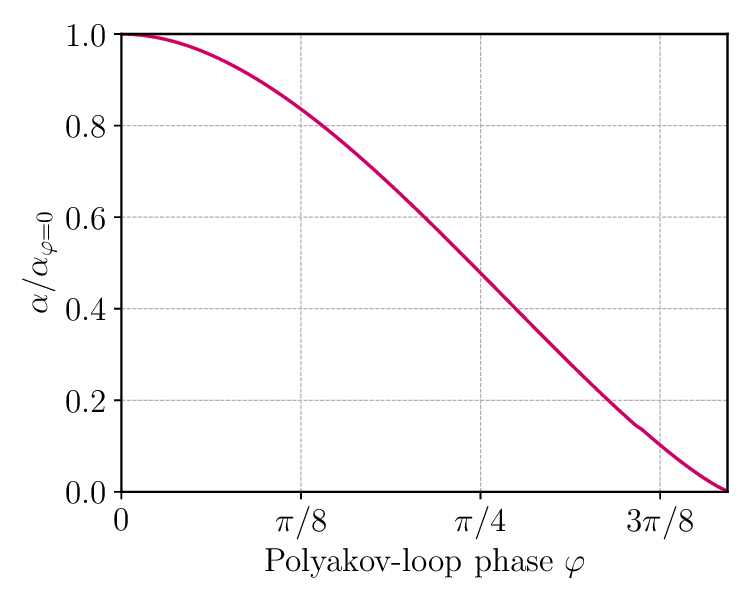}%
  \caption{%
    First-order phase-transition lines in the plane of the Polyakov-loop phase $\varphi$
    for
    the critical temperature $\Tc(\varphi)$ (left) and
    the transition strength $\alpha(\varphi)$ (right)
    normalized to their trivial-holonomy
    values
    $T_{\rmii{c},\varphi=0}$ and
    $\alpha_{\varphi=0}$.
    The Polyakov loop phase $\varphi$ is varied at
    benchmark point~\eqref{eq:BM1},
    between $\varphi=0$ (trivial holonomy) and
    $\varphi_\rmi{max}\approx1.32$,
    after which the transition becomes second order and $\alpha$ vanishes.
    A non-trivial holonomy
    increases $\Tc$ and
    lowers $\alpha$ compared to the trivial $\varphi=0$.
  }
  \label{fig:Tc:polyakov}
\end{figure}
Figure~\ref{fig:Tc:polyakov} shows that turning on the holonomy
monotonically increases $\Tc$ and decreases $\alpha$
relative to the trivial-holonomy case,
until the second-order endpoint $\varphi_\rmi{max} \approx 1.32$
where $\alpha$ vanishes.
The diagrammatic estimate sits at $\varphi=0$ and therefore
underpredicts $\Tc$ and overpredicts the transition strength $\alpha$,
whereas the heat kernel result captures the full Polyakov loop dependence,
as discussed in sec.~\ref{sec:dr:differences}.

\section{Comparison with diagrammatic dimensional reduction}
\label{sec:dr}

We now compare
the operator basis and matching coefficients
obtained from the heat-kernel construction in
sec.~\ref{sec:AH:heatKernel}
with the results of diagrammatic dimensional reduction.
The most directly comparable computation is the NNLO matching of the
Abelian Higgs model in~\cite{Bernardo:2025vkz},
whose off-shell soft-scale basis is reproduced in
tab.~\ref{tab:dr:dim6}.
A diagrammatic dimension-six matching of a real-scalar Yukawa model has been
carried out in~\cite{Chala:2024xll}.
A general-model automation that systematizes the construction of
the dimensionally reduced higher-dimensional operator basis
has recently been provided
in~\cite{Bernardo:2026nyq} and~\cite{Fuentes-Martin:2026bhr}.

\subsection{Diagrammatic operator basis}
\label{sec:dr:basis}

In the soft-scale three-dimensional EFT of
the Abelian Higgs model,
ref.~\cite{Bernardo:2025vkz} introduces
a redundant off-shell basis of
dimension-six operators built
using the background field gauge~\cite{Abbott:1980hw}.
The constituting fields are
the background complex scalar $\bphi$,
the background spatial gauge field $\bA_i$ with field strength
$\bG_{ij}$,
and the background temporal scalar $\bA_0$.
The physical (on-shell) operators
together with the operators that are redundant by
the scalar and gauge equations of motion are summarized in
tab.~\ref{tab:dr:dim6}.
\begin{table}[t]
  \centering
  \renewcommand{\arraystretch}{1.1}
  \begin{tabular}[t]{|l|l|}
    \hline
    \multicolumn{2}{|c|}{Physical dimension-6 operators}\\
    \hline\hline
    $\mathcal{O}^{\rmii{DR}}_1$ & $\bG_{ij}\bG_{ij}\,\bA_0^2$\\\hline
    $\mathcal{O}^{\rmii{DR}}_2$ & $\bG_{ij}\bG_{ij}\,\bphi^\dagger\bphi$\\\hline
    $\mathcal{O}^{\rmii{DR}}_3$ & $(\bD_i\bphi^\dagger \bD_i\bphi)\,(\bphi^\dagger\bphi)$\\\hline
    $\mathcal{O}^{\rmii{DR}}_4$ & $(\bD_i\bphi^\dagger \bD_i\bphi)\,\bA_0^2$\\\hline
    $\mathcal{O}^{\rmii{DR}}_5$ & $\bA_0^6$\\\hline
    $\mathcal{O}^{\rmii{DR}}_6$ & $\bA_0^4\,\bphi^\dagger\bphi$\\\hline
    $\mathcal{O}^{\rmii{DR}}_7$ & $\bA_0^2\,(\bphi^\dagger\bphi)^2$\\\hline
    $\mathcal{O}^{\rmii{DR}}_8$ & $(\bphi^\dagger\bphi)^3$\\\hline
  \end{tabular}
  \hspace{0.8cm}
  \begin{tabular}[t]{|l|l|}
    \hline
    \multicolumn{2}{|c|}{Redundant operators}\\
    \hline\hline
    $\mathcal{R}^{\rmii{DR}}_1$ & $(\partial_i \bG_{ij})^2$\\\hline
    $\mathcal{R}^{\rmii{DR}}_2$ & $\bA_0\,\Box^2 \bA_0$\\\hline
    $\mathcal{R}^{\rmii{DR}}_3$ & $\bA_0^3\,\Box \bA_0$\\\hline
    $\mathcal{R}^{\rmii{DR}}_4$ & $(\bD^2\bphi^\dagger)(\bD^2\bphi)$\\\hline
    $\mathcal{R}^{\rmii{DR}}_5$ & $(\bphi^\dagger\bphi)\bigl(\bphi^\dagger \bD^2\bphi+\mathrm{h.c.}\bigr)$\\\hline
    $\mathcal{R}^{\rmii{DR}}_6$ & $(\partial_i \bG_{ij})\,i\bphi^\dagger(\bD_j\bphi)$\\\hline
    $\mathcal{R}^{\rmii{DR}}_7$ & $(\bphi^\dagger\bphi)\,\bA_0\Box \bA_0$\\\hline
    $\mathcal{R}^{\rmii{DR}}_8$ & $\bphi^\dagger(\bD_i^2\bphi)\bA_0^2 + \mathrm{h.c.}$\\\hline
  \end{tabular}
  \caption{%
    Off-shell dimension-six operator basis of
    non-redundant ($\mathcal{O}$) and redundant ($\mathcal{R}$) operators
    in the Abelian Higgs model in the soft
    dimensionally-reduced (DR) three-dimensional EFT,
    as constructed diagrammatically in~\cite{Bernardo:2025vkz}
    using the background field gauge~\cite{Abbott:1980hw}.
    The operators $\mathcal{R}$ are redundant by the
    scalar and gauge equations of motion.
  }
  \label{tab:dr:dim6}
\end{table}

\subsection{Heat kernel basis and dictionary}
\label{sec:dr:dictionary}

The heat kernel result for $\mathcal{L}_{\rmi{eff}}$
in eq.~\eqref{eq:AH:heatKernel:Leff} is naturally organized
in terms of background-covariant building blocks
$\bphi$, $\bD_i$,
the field strength $\bG_{ij}$,
the electric components $E_i=-[\bD_i,\bD_0]$,
and the current $\bJ_i\sim(\bD_j \bG_{ij})$.
The Polyakov loop dependence is resummed into the
finite-temperature master sums $S_\Omega$ and $I_\Omega$,
so that no explicit tower of $\bA_0^n$ operators appears.
A schematic dictionary between the heat kernel operators
and the diagrammatic basis of tab.~\ref{tab:dr:dim6}
is given in tab.~\ref{tab:hk:dr:dictionary}.
\begin{table}[t]
  \centering
  \renewcommand{\arraystretch}{1.2}
  \begin{tabular}{|c|c||c|}
    \hline
    Heat kernel operators
      & Diagrammatic operators
      & Comment
    \\\hline\hline
    $\bphi^2$, $\bphi^4$, $\bphi^6$
      &
      $\phi^\dagger\phi$,
      $(\phi^\dagger\phi)^2$,
      $\mathcal{O}^{\rmii{DR}}_8$
      & potential, on-shell
    \\\hline
    $(\bD_i\bphi_a)(\bD_i\bphi_a)$
      & $(D_i\phi)^\dagger(D_i\phi)$
      & kinetic, on-shell
    \\\hline
    $(\bD_i\bphi^2)^2$
      & $\mathcal{O}^{\rmii{DR}}_3$
      & via integration by parts
    \\\hline
    $\bG_{ij}\bG_{ij}$
      & $\bG_{ij}\bG_{ij}$
      & gauge kinetic
    \\\hline
    $\bphi^2\,\bG_{ij}\bG_{ij}$
      & $\mathcal{O}^{\rmii{DR}}_2$
      & on-shell
    \\\hline
        \multirow{2}{*}{$\bJ_i^2 = (\bD_j \bG_{ij})^2$}
      & \multirow{2}{*}{$\mathcal{R}^{\rmii{DR}}_1$}
      & EOM-redundant, traded via
    \\
      & & $\bD_j \bG_{ij}\sim \bphi^\dagger\bD_i\bphi+\mathrm{h.c.}$
    \\\hline
    $E_i^2$,
    $E_{i0}^2$,
    $E_{i00}^{ }E_i^{ }$,
    $E_{ii0}^{ }$
      & $\subset
        \mathcal{O}^{\rmii{DR}}_{1,4}$,
      $\mathcal{R}^{\rmii{DR}}_{2,7,8}$
      & resummed Polyakov dressing
    \\\hline
    $\bA_0^{2k}\bphi^{2m}$
      & $\subset
        \mathcal{O}^{\rmii{DR}}_{4,6,7}$,
        $\mathcal{R}^{\rmii{DR}}_{7,8}$
      & in static limit 
    \\\hline
         \multirow{2}{*}{ (no explicit $\bA_0^n$)}
      & \multirow{2}{*}{%
        $\mathcal{O}^{\rmii{DR}}_{5}$,
        $\mathcal{R}^{\rmii{DR}}_{2,3}$
        }
      & resummed via gauge-space\\
      && averaged $\Omega$
      in $I_\Omega,S_\Omega$
      in eq.~\eqref{eq:heatKernel:IOmega}
    \\\hline
  \end{tabular}
  \caption{%
    Schematic correspondence between
    the heat-kernel operators that appear in
    eq.~\eqref{eq:AH:heatKernel:Leff} and
    the diagrammatic operator basis of
    tab.~\ref{tab:dr:dim6}.
    The covariant derivative
    $\bD_\mu = \partial_\mu + \bA_\mu$
    contains the background gauge field, and
    $\bG_{ij}\equiv [\bD_i,\bD_j]$
    is the scalar-sector field strength.
  }
  \label{tab:hk:dr:dictionary}
\end{table}

\subsection{Structural differences}
\label{sec:dr:differences}

The two constructions are
structurally different,
and we now discuss the main differences and how they are reconciled:
\begin{itemize}
  \item[(i)]
  \emph{Full Polyakov loop treatment.}
  One shortcoming of
  the diagrammatic basis of~\cite{Bernardo:2025vkz}
  is the explicit emergence of towers
  $\bA_0^n$
  (operators
  $\mathcal{O}^{\rmii{DR}}_{5}$,
  $\mathcal{R}^{\rmii{DR}}_{2,3}$).
  In the heat kernel,
  only operators of the type
  $\bA_0^{2k}(\bphi^\dagger\bphi)^m$
  ($\mathcal{O}^{\rmii{DR}}_{4,6,7}$,
  $\mathcal{R}^{\rmii{DR}}_{7,8}$
  )
  appear in the static limit
  where $(Q\bphi_a)=\bA_0 \bphi_a $.
  Pure $\bA_0^n$ towers could be
  generated by a local expansion of the Polyakov loop holonomy
  $\Omega = e^{-\beta \bA_0}$ from eq.~\eqref{eq:polyakov}.
  Such an expansion was already constructed
  in QCD~\cite{Chapman:1994vk,Megias:2003ui,Bernardo:2026nyq,romainguillermo2026} and
  recently discussed in a general-model treatment
  in~\cite{Chakrabortty:2026swu,Bernardo:2026nyq}.

  By construction, pure $\bA_0$ contributions are
  contained in the master Matsubara sums
  $I_\Omega$ and $S_\Omega$
  (cf.\ appendix~\ref{sec:masterIntegrals})
  via the full Polyakov loop and
  its gauge space averaged value.
  See eq.~\eqref{eq:heatKernel:IOmega:def} and below for a definition.
  This corresponds to a resummation of the Polyakov loop effects 
  from the perspective of the diagrammatic construction.
  Conversely, the heat kernel basis is a generalization
  of the diagrammatic approach and the Polyakov loop can directly affect phase-transition
  thermodynamics
  as discussed in sec.~\ref{sec:PolyakovLoop:effects}.

  \item[(ii)]
  \emph{Redundant versus non-redundant operator basis.}
  The effective action computed using the heat kernel method contains
  redundant operators, as in the process, the field redefinition, equation of motion, and IBP are not employed. Operators such as $\bJ_i^2=(\bD_j \bG_{ij})^2$
  and the higher-derivative $E$-structures
  correspond to entries in the redundant column of
  tab.~\ref{tab:dr:dim6}, and
  are eliminated by the field redefinitions detailed
  in~\cite{Bernardo:2025vkz,Bernardo:2026nyq}.
  Concretely, the gauge EOM
  $\bD_j \bG_{ij}\sim \bigl(\bphi^\dagger\bD_i\bphi-(\bD_i\bphi)^\dagger\bphi\bigr)$
  trades $\bJ_i^2$ together with the EOM for the background scalars $\bD_i^2\hat{\phi} \sim \lambda \hat{\phi}^3$ for combinations of
  $(\bD_i\bphi^\dagger \bD_i\bphi)(\bphi^\dagger\bphi)$
  and the scalar potential operators. 

  \item[(iii)]
  \emph{Algebraic versus diagrammatic organization.} 
  The heat kernel coefficients group operators by
  background-covariant building blocks
  ($U$, $\bG_{ij}$, $E_i$, $\bJ_i$),
  while the diagrammatic construction of~\cite{Chala:2024xll,Bernardo:2025vkz}
  enumerates Lorentz- and gauge-invariant local operators
  to a given mass dimension.
  Both approaches yield the same on-shell physics
  once the EOM redundancies are removed and
  the Polyakov loop is locally expanded.
  Such an expansion, however, yields
  an incomplete inclusion of
  the full Polyakov loop.

\end{itemize}
A complete operator-by-operator matching of
Wilson coefficients in the non-redundant diagrammatic operator basis
is listed in~\cite{Bernardo:2026nyq} for the Abelian Higgs model
and automated for general models in~\cite{Fuentes-Martin:2026bhr}.
In comparison to the heat kernel result, these diagrammatic automations lack
the complete Polyakov loop dependence.
We leave such automation of functional-matching in the heat kernel method for future work.

\section{Conclusions and outlook}
\label{sec:outlook}

In this work, we constructed
the finite-temperature one-loop effective action
in the thermal EFT of the Abelian Higgs model
up to dimension six by applying the heat kernel method.
We demonstrated two self-consistent methods
to compute the 3D static thermal effective action:
(i) directly integrating out at finite temperature, and
(ii) generating the thermal action from the zero-temperature result through matching.
The resulting operator basis is organized in terms of
background-covariant building blocks
($\bphi$, $\bD_i$, $\bG_{ij}$, $E_i$, $\bJ_i$)
together with the temporal holonomy $\Omega$, and after field redefinitions
it reproduces, on shell, the diagrammatic basis
of~\cite{Chala:2024xll,Bernardo:2025vkz} and the {\tt DRalgo}
output of~\cite{Bernardo:2026nyq}.

A central feature of the heat kernel construction is its
robust treatment of the Polyakov loop.
Its full holonomy dependence is resummed into
the thermal master integrals through a gauge-space averaging,
without expanding around trivial holonomy.
This resummation is absent in
the diagrammatic approach, which instead generates
a tower of temporal gauge-field operators of the form
$\bA_0^{2k}$,
corresponding to a local expansion of $\Omega$.
The Polyakov loop can be treated as an order parameter
whose traceless condition fixes the value of $\beta\bA_0$
in terms of
the gauge charge of the infrared degrees of freedom.
We demonstrated that the Polyakov loop has a significant impact on
the phase transition,
increasing the critical temperature $\Tc$ and
decreasing the transition strength $\alpha$ relative to the trivial-holonomy case.
This has direct implications for the corresponding gravitational-wave signal,
and a systematic investigation of the Polyakov-loop effects on
the gravitational-wave spectrum is left for future work.

The methodology developed here is not specific to the Abelian Higgs
model and extends naturally to
non-Abelian gauge theories~\cite{%
  Banerjee:2023iiv,Banerjee:2023xak,Chakrabortty:2023yke,Balui:2026ghs} and
to the SMEFT,
based on~\cite{Chakrabortty:2026swu}.
We note that our second method from sec.~\ref{sec:AH:heatKernel:zeroT} can be directly applied to
construct thermal effective operators from
their zero-temperature counterparts~\cite{Banerjee:2023iiv,Banerjee:2023xak,Chakrabortty:2023yke}
through matching.

The results of this article open
promising directions for automating the functional matching of
the heat kernel at finite temperature, based on
existing zero- and finite-temperature tools~\cite{%
  DasBakshi:2018vni,Fuentes-Martin:2022jrf,Fuentes-Martin:2026bhr},
and including fermionic loops~\cite{Banerjee:2023xak,Chakrabortty:2024wto}.
Another direction is to apply
the heat-kernel effective action
to bubble nucleation and to investigate
the impact of the Polyakov loop on the nucleation rate and
the gravitational-wave signal
across the parameter space relevant for LISA.
We leave these developments for future work.

\section*{Acknowledgments}

We thank
Fabio Bernardo,
Romain Guillermo Reinle,
Tuomas V.I. Tenkanen, and
Jorinde van de Vis
for illuminating discussions.
JC acknowledges the hospitality of HRI, Allahabad, India, where part of the research was done.
SB, JC, DD, and T acknowledge support from the Science and Engineering Research Board (SERB), Government of India, under the Project SERB/PHY/2023799.
PS was supported by
the Swiss National Science Foundation (SNSF) under grant
\href{https://data.snf.ch/grants/grant/215997}{\tt PZ00P2-215997}.

\appendix
\renewcommand{\thesection}{\Alph{section}}
\renewcommand{\thesubsection}{\Alph{section}.\arabic{subsection}}
\renewcommand{\theequation}{\Alph{section}.\arabic{equation}}

\section{Master integrals and sums}
\label{sec:masterIntegrals}

In this section,
we define the master integrals and sums that appear in the local
effective Lagrangian.
To this end, we introduce the thermal
wave-functions
\begin{align}
    \varphi_k(\Omega; t/\beta^2)
    = (4\pi t)^{1/2} \, \frac{1}{\beta} \sum_{p_0} t^{k/2} \, Q^k \, e^{Q^2 t}
    \,,
\end{align}
where
$Q =
\bD_0 + i p_0 \equiv
\bigl[\frac{2n\pi i}{\beta} -\frac{\ln \Omega}{\beta}\bigr]$,
$n\in \mathbb{Z}$ and
$\Omega=e^{-\beta \bA_0}$ is
the Polyakov loop as defined in eq.~\eqref{eq:polyakov}.
The contribution from the Polyakov loop is captured in the Matsubara sums and 
the corresponding one-loop master
integrals are
\begin{align}
  \label{eq:heatKernel:IOmega:def}
    I_{\Omega}^m(k;l) &=
       \tmu^{2\epsilon}\int_0^\infty \frac{{\rm d}t}{t} \frac{ e^{-m^2 t}}{(4\pi t)^{\frac{d+1}{2}}} \, t^l \, \varphi_k(\Omega;t/\beta^2)
       =
      \frac{ \tmu^{2\epsilon} }{\beta} 
      \int_0^\infty \frac{{\rm d}t}{t}
      \frac{e^{-m^2 t}}{(4\pi t)^{\frac{d}{2}}} t^l
      \sum_{p_0} t^{k/2} \, Q^k \, e^{Q^2 t}
    \nn[1mm] &=
    \frac{ \tmu^{2\epsilon}}{\beta (4\pi)^{\frac{d}{2}}}
    \Bigl( \frac{2\pi i}{\beta} \Bigr)^k
    \sum_{n}\,
      [n+\tilde{n}]^k
    \nn &
    \qquad \times
      \int_{0}^{\infty}\!{\rm d}t\;
      t^{\frac{2l+k-d-2}{2}}
      \exp\Bigl\{
        -\Bigl[
          m^2
         + \Bigl(\frac{2\pi}{\beta}\Bigr)^2 (n+\tilde{n})^2
        \Bigr] t
      \Bigr\}
  \,,
\end{align}
where $\tmu^2 = 4\pi e^{-\gammaE} \mu^2$ is the
$\overline{\text{MS}}$ renormalization scale, and
$\tilde{n} = \frac{i}{2\pi} \langle\ln \Omega \rangle \in \mathbb{R}$
where the average is taken over the gauge states
(cf.\ eq.~\eqref{eq:polyakov:shift}).
Performing the proper-time integral yields the master integral as a
sum over Matsubara modes,
\begin{align}
  \label{eq:heatKernel:IOmega}
    I_{\Omega}^m(k;l) &=
      \frac{ \tmu^{2\epsilon}}{\beta (4\pi)^{\frac{d}{2}}}  
      \Bigl( \frac{2\pi i}{\beta} \Bigr)^k
      \sum_{n}
      \,[n+\tilde{n}]^k
    \nn &
    \qquad
    \times
      \Bigl[ m^2 + \Bigl(\frac{2\pi}{\beta}\Bigr)^2 (n+\tilde{n})^2 \Bigr]^{-\frac{2l+k-d}{2}}
      \Gamma\Bigl( \frac{2l + k - d}{2} \Bigr)
    \,.
\end{align}

The integral in eq.~\eqref{eq:heatKernel:IOmega} is the master integral for
the one-loop effective Lagrangian.
It contains both the
zero-temperature vacuum contribution and the finite-temperature
thermal corrections, which can be separated.
For $k=0$,
this separation is given by 
\begin{align}
  \label{eq:heatKernel:IOmega:split}
  I_{\Omega}^m(0;l)
  &= \underbrace{
       I_{\rmi{vac}}^m(l)
     }_{\displaystyle \text{($T=0$)}}
   + \underbrace{
      S_{\Omega}^m(0;l)
     }_{\displaystyle \text{($T\neq 0$)}}
  \,.
\end{align}
The first part,
the $T$-independent term,
corresponds to the zero-temperature vacuum contribution,
while
the second part captures the thermal effects.
They are given by
\begin{align}
  I_{\rmi{vac}}^m(l) &=
   \tmu^{2\epsilon}
  \frac{[m^2]^{\frac{D}{2} - l}}{(4\pi)^{\frac{D}{2}}} \Gamma\Bigl(l - \frac{D}{2}\Bigr)
  \,,\\[1mm]
  \label{eq:heatKernel:SOmega:compact}
    S_{\Omega}^m(0;l) &=
    \frac{4 \tmu^{2\epsilon} }{(4\pi)^\frac{D}{2}}
    \Bigl( \frac{\beta}{2m} \Bigr)^{l-\frac{D}{2}}
    \sum_{n=1}^{\infty}
      n^{l-\frac{D}{2}} \cos(2\pi n \tilde{n})\,
      \mathbb{K}_{l-\frac{D}{2}}(n m \beta)
  \,,
\end{align}
where
the thermal sum $S_{\Omega}^m(0;l)$ is UV finite and
$\mathbb{K}_\nu$ is the modified Bessel function of the second kind.

Some special cases of the $I_\Omega$
functions are~\cite{Chakrabortty:2024wto}%
\footnote{%
  We label the functions with the corresponding mass parameters explicitly.
}
\begin{align}
  I_{\Omega}^m(0;0) &=
  \frac{m^4}{32\pi^2}
  \Bigl(
      \ln\frac{ \tmu^2}{m^2}
    + \frac{3}{2}
  \Bigr) + S^m_{\Omega}(0;0)
  \,,\\[1mm]
  I_{\Omega}^m(0;1) &= -\frac{m^2}{(4\pi)^2}
  \Bigl(
     \ln\frac{ \tmu^2}{m^2}
    + 1
  \Bigr) + S^m_{\Omega}(0;1)
  \,,\\[1mm]
  I^m_{\Omega}(0;2) &= \frac{1}{(4\pi)^2} \Bigl(
      \ln\frac{ \tmu^2}{m^2}
    \Bigr)
  + S^m_{\Omega}(0;2)
  \,,\\[1mm]
  I_{\Omega}^m(0;3) &=
      \frac{1}{(4\pi)^2 m^2}
    + S^m_{\Omega}(0;3)
  \,.
\end{align}
The thermal sums $S^m_{\Omega}(0;l)$, for $l \in \mathbb{Z}_{\geq 0}$,
are derived in~\cite{Meisinger:2001fi}.
Introducing the Polyakov loop phase
$\varphi \equiv 2\pi\tilde{n}\hspace{0.01cm} \bmod 2\pi \in [0,2\pi)$,
they read \cite{Meisinger:2001fi}
\begin{align}
  S^m_{\Omega}(0;0) &=
  \frac{m^2}{\pi^2 \beta^2} \sum_{n=1}^\infty \frac{1}{n^2} \cos(2\pi n \tilde{n})\,\mathbb{K}_{-2}(n m \beta)
  \nn &=
  \frac{m^2}{\pi^2 \beta^2} \bigg\{
      \frac{m^2 \beta^2}{16} \bigg[ \ln\frac{m\beta}{4\pi} + \gammaE - \frac{3}{4} \bigg]
    - \frac{1}{2} \bigg[ \frac{1}{4} \varphi^2 - \frac{\pi}{2} \varphi + \frac{\pi^2}{6} \bigg]
  \nn &
  \hphantom{{}\frac{m^2}{\pi^2 \beta^2} \bigg\{}
    + \frac{2}{m^2 \beta^2} \bigg[ - \frac{1}{48} \varphi^4 + \frac{\pi}{12} \varphi^3 - \frac{\pi^2}{12} \varphi^2 + \frac{\pi^4}{90} \bigg]
  \nn &
  \hphantom{{}\frac{m^2}{\pi^2 \beta^2} \bigg\{}
    + \frac{\pi}{2 m^2 \beta^2} \sum_{\ell \in \mathbb{Z},\, \ell \neq 0} \bigg[ \frac{1}{3} \Big[ m^2 \beta^2 + 4\pi^2(\tilde{n} - \ell)^2 \Big]^{3/2} - \frac{8\pi^3}{3} |\tilde{n} - \ell|^3
  \nn &
  \hphantom{{}\frac{m^2}{\pi^2 \beta^2} \bigg\{+\frac{\pi}{2 m^2 \beta^2} \sum_{\ell \in \mathbb{Z},\, \ell \neq 0} \bigg\{}
  - \pi m^2 \beta^2 |\tilde{n} - \ell| - \frac{m^4 \beta^4}{16\pi|\ell|} \bigg]
  \bigg\}
  \,,
  \\[2mm]
  S^m_{\Omega}(0;1) &=
      \frac{m}{2\pi^2 \beta}\sum_{n=1}^{\infty} \frac{1}{n} \cos(2\pi n\tilde{n})\,\mathbb{K}_1(n m \beta)
      \nn & =
      \frac{m}{2\pi^2 \beta}\bigg\{
        - \frac{1}{4} m\beta \bigg[ \ln\frac{m\beta}{4\pi} + \gammaE - \frac{1}{2} \bigg]
        + \frac{1}{m\beta} \bigg[ \frac{1}{4}\varphi^2 - \frac{\pi}{2}\varphi + \frac{\pi^2}{6} \bigg]
      \nn &
      \hphantom{{}=\frac{m}{2\pi^2 \beta}\bigg\{}
      - \frac{\pi}{2 m\beta} \sum_{ \ell \, \in  \mathbb{Z}, \, \ell \neq 0} \bigg[
            \sqrt{m^2 \beta^2 + 4\pi^2(\tilde{n} - \ell)^2}
          - 2\pi|\tilde{n} - \ell|
          - \frac{m^2 \beta^2}{4\pi|\ell|}
          \bigg]
      \bigg\}
  \,,
  \\[2mm]
  S^m_{\Omega}(0;2) &=
    \frac{1}{(2\pi)^2} \sum_{n=1}^{\infty} \cos(2\pi n \tilde{n}) \mathbb{K}_0(n m \beta)
    \nn &=
    \frac{1}{(2\pi)^2} \bigg\{
        \frac{1}{2} \Big[ \gammaE + \ln\frac{m\beta}{4\pi} \Big]
      + \frac{\pi}{2} \sum_{ \ell \, \in  \mathbb{Z}, \, \ell \neq 0} \bigg[
          \frac{1}{\sqrt{m^2\beta^2 + 4\pi^2(\tilde{n} - \ell)^2}}
        - \frac{1}{2\pi|\ell|} \bigg]
      \bigg\}
    \,.
\end{align}

The split in eq.~\eqref{eq:heatKernel:IOmega:split}
can also be made manifest by
Poisson resummation of the Matsubara sum
in eq.~\eqref{eq:heatKernel:IOmega},
\begin{align}
  \label{eq:poissonResummation}
  \frac{1}{\beta} \sum_{n\in\mathbb{Z}} f\bigl(\tfrac{2\pi}{\beta}(n+\tilde n)\bigr)
  = \sum_{w\in\mathbb{Z}} e^{2\pi i w \tilde n}\,
  \int\!\frac{{\rm d} p_0}{2\pi}\,e^{i p_0 w\beta} f(p_0)
  \,.
\end{align}
After separating the $w=0$ contribution from the rest
and reinstating
the $d$-dimensional spatial integration,%
\footnote{
  The $(4\pi)^{d/2}$ prefactor in \eqref{eq:heatKernel:IOmega} already
  encodes the $d$-dimensional spatial Gaussian integration
  $\int_\vec{p} e^{-p^2 t} = (4\pi t)^{-d/2}$
  that was performed via the Schwinger parametrization.
}
the master integral splits
into a vacuum and thermal part
as in eq.~\eqref{eq:heatKernel:IOmega:split}.
The $w=0$ piece is the standard $D$-dimensional vacuum integral,
\begin{align}
  I_{\rmi{vac}}^m(0) &=
  -\int_P \ln(P^2 + m^2)
  =  \tmu^{2\epsilon}
  \frac{[m^2]^{\frac{D}{2}}}{(4\pi)^{\frac{D}{2}}} \frac{\Gamma\bigl(-\frac{D}{2}\bigr)}{\Gamma(1)}
  \,,\\[1mm]
  I_{\rmi{vac}}^m(l) &=
  \Gamma(l)\int_P \frac{1}{[P^2 + m^2]^l}
  =  \tmu^{2\epsilon}
  \frac{[m^2]^{\frac{D}{2} - l}}{(4\pi)^{\frac{D}{2}}} \Gamma\Bigl(l - \frac{D}{2}\Bigr)
  \,.
\end{align}
In dimensional
regularization with $D = 4 - 2\epsilon$, the poles in $\epsilon$ of
these vacuum integrals are absorbed into the
$\overline{\text{MS}}$ counterterms of the underlying theory.
Conversely, the $w\neq 0$ contributions are exponentially suppressed
at large $|p_0|$ and are therefore UV finite.
Using the symmetry of
the sum in eq.~\eqref{eq:poissonResummation} and
identifying the $p_0$
integral as a modified Bessel function of the second kind $\mathbb{K}_\nu$,
the thermal part can be evaluated
via the closed-form sum in $d=3$
in eq.~\eqref{eq:heatKernel:SOmega:compact}.

\section{Heat kernel coefficients}
\label{sec:heatKernel coefficients}

\subsection{%
  Thermal heat kernel coefficients
  from zero temperature coefficients}
\label{app:Z-THKC}

In this section,
we present the thermal heat kernel coefficients
obtained from matching the zero temperature heat kernel coefficients
to the thermal heat kernel coefficients
following the procedure described in
sec.~\ref{sec:AH:heatKernel:zeroT} and~\cite{Chakrabortty:2024wto}.

We match for different values of
$k$ starting from 0 in different powers of $t$, 
which gives the thermal heat kernel coefficients as
\begin{align}
  \tilde{f}^\rmii{$S$}_0 &= 2
  \,,\\
  \tilde{f}^\rmii{$SG$}_0 &= -\frac{2g^2}{(\Delta_{12}^2)^3} ([\bD_i ,(\bD_\mu \bphi_a)] [\bD_i, (\bD_\mu\bphi_a)])
  \,,\\
  \tilde{f}^\rmii{$G$}_0 &= -\eta_{\mu\mu}
  \,,\\
  \tilde{f}^\rmii{$GS$}_0 &= \frac{2g^2}{(\Delta_{12}^2)^3} ([\bD_i ,(\bD_\mu \bphi_a)] [\bD_i, (\bD_\mu\bphi_a)])
  \,,
\end{align}
where the superscript
$S$ and $G$ represent the scalar and gauge sectors, respectively, and the subscript represents the power of $t$.
Here, we make use of the compact notation,
\begin{equation}
  [\bD_i ,(\bD_\mu \bphi_a)] [\bD_i, (\bD_\mu\bphi_a)]=
    [\bD_i ,(Q \bphi_a)] [\bD_i, (Q \bphi_a)]
  + [\bD_i ,(\bD_j \bphi_a)] [\bD_i, (\bD_j\bphi_a)]
  \,.
\end{equation}
In general, we make the replacement
\begin{equation}
   ...\,\bD_\mu \bphi\,...\,\bD_\mu \bphi\,... \rightarrow
   ...\,Q \bphi\,...\,Q \bphi\,...\,+...\,\bD_j \bphi\,...\,\bD_j \bphi\,...
   \,,
\end{equation}
where the ellipsis represents insertion of other operators and/or
commutators of derivatives.
Using our compact notation,
we write the higher-order thermal heat kernel coefficients,
starting at $\mathcal{O}(t)$,
\begin{align}
  \tilde{f}^\rmii{$S$}_1 &= U_\rmii{$S$}
  \,,\\
  \tilde{f}^\rmii{$SG$}_1 &=
    \frac{g^2}{\Delta^2_{12}}(\bD_\mu \bphi_a)( \bD_\mu \bphi_a)+ \frac{g^2}{(\Delta_{12}^2)^2} \left[ (\bD_\mu \bphi_a)( \bD_\mu \bphi_b) U_{ab} - (\bD_\mu \bphi_a)(\bD_\mu \bphi_a) U_s \right]
  \nn &
  + \frac{g^2}{(\Delta^2_{12})^2}(\bD_\mu \bphi_a)U_{\mu\nu}(\widehat{D }_{\nu}\bphi_a)+\frac{g^2}{(\Delta^2_{12})^2}[\bD_i , (\bD_\nu \bphi_a)][\bD_i , (\bD_\nu\bphi_a)]
  \,,\\[2mm]
  \tilde{f}^\rmii{$G$}_1 &= U_\rmii{$G$}
  \,,\\
  \tilde{f}^\rmii{$GS$}_1 &=
  -\frac{g^2}{\Delta^2_{12}}(\bD_\mu \bphi_a)( \bD_\mu \bphi_a)- \frac{g^2}{(\Delta_{12}^2)^2} \left[ (\bD_\mu \bphi_a)( \bD_\mu \bphi_b) U_{ab} - (\bD_\mu \bphi_a)(\bD_\mu \bphi_a) U_s \right]
  \nn &
  - \frac{g^2}{(\Delta^2_{12})^2}(\bD_\mu \bphi_a)U_{\mu\nu}(\widehat{D }_{\nu}\bphi_a)+\frac{g^2}{(\Delta^2_{12})^2}[\bD_i , (\bD_\nu \bphi_a)][\bD_i , (\bD_\nu\bphi_a)]
  \,,
\end{align}
for the gauge and scalar sectors.
At $\mathcal{O}(t^2)$, the thermal heat kernel coefficients are
\begin{align}
  \tilde{f}^\rmii{$S$}_2 &=
        U_{ab}U_{ba}
      - \frac{1}{3}[\bD_i, [\bD_i, U_\rmii{$S$}]]
      + 2\frac{[\bD_i, [\bD_i, Q^2]]}{3}
      + 2 \frac{\bG_{ij} \bG_{ij}}{6}
  \,,\\
  \tilde{f}^\rmii{$SG$}_2 &=
      \frac{2g^2}{(\Delta^2_{12})}[(\bD_\mu \bphi_a)(\bD_\mu\bphi_a)U_\rmii{$S$}-(\bD_\mu \bphi_a)( \bD_\mu \bphi_b) U_{ab}]
    - \frac{g^2}{\Delta^2_{12}}\frac{[\bD_i,[\bD_i , (\bD_\mu \bphi_a)(\bD_\mu \bphi_a)]]}{3}
  \,,\\
  \tilde{f}^\rmii{$G$}_2 &=
      U_{\mu\nu}U_{\nu\mu}
    - \frac{1}{3}[\bD_i, [\bD_i, U_\rmii{$G$}]]
    - \eta_{\mu\mu}\frac{[\bD_i, [\bD_i, Q^2]]}{3}
    - \eta_{\mu\mu} \frac{\bG_{ij} \bG_{ij}}{6}
  \,,\\
  \tilde{f}^\rmii{$GS$}_2 &=
    - \frac{2g^2}{(\Delta^2_{12})}[(\bD_\mu \bphi_a)(\bD_\nu\bphi_a)U_{\mu\nu}]
    - \frac{g^2}{\Delta^2_{12}}\frac{[\bD_i,[\bD_i , (\bD_\mu \bphi_a)(\bD_\mu \bphi_a)]]}{3}
  \,,
\end{align}
where
\begin{align}
  \label{eq:commutatorCancellations}
  [Q^2, U_\rmii{$S$}] =
  [Q^2, U_\rmii{$G$}] =
  [Q^2, U_{ab}] = 
  [Q^2, U_{\mu\nu}] &= 0
  \,,
\end{align}
in the static limit and for Abelian gauge fields.

The thermal heat kernel coefficients
at $\mathcal{O}(t^3)$ are more involved, and
we write them directly having taken into account the cancellations
of eq.~\eqref{eq:commutatorCancellations}.
For the scalar sector,
we find
\begin{align}
  \tilde{f}^\rmii{$S$}_3 &=  
    U_{ab}U_{bc}U_{ca}
  - Q^2 U_{\rmii{$S$};ii}
  + 2 Q^2 (Q^2)_{;ii}
  + \frac{1}{2}U_\rmii{$S$}\bG_{ij}\bG_{ij}
  + \frac{1}{10}U_{\rmii{$S$};iijj} -\frac{1}{5}(Q^2)_{;iijj}
  \nn &
  - \frac{2}{10}(\bJ_i)^2
  + \frac{2}{15}\bG_{ij}\bG_{jk}\bG_{ki}
  - \frac{2}{30}[\bD_i,[\bD_j,[\bD_j,\bJ_i]]]
  \nn &
  - \frac{1}{2}\Bigl[
      [\bD_i,[\bD_i,U_{ab}U_{ba}]]
    - (Q^2U_\rmii{$S$})_{ii}
    - (U_\rmii{$S$} Q^2)_{ii}
    + 2(Q^4)_{ii}
  \Bigr]
  \nn &
  + \frac{1}{2}\Bigl[
      [\bD_i,U_{ab}][\bD_i,U_{ba}]
    -( Q^2)_{;i}U_{\rmii{$S$};i}
    - U_{\rmii{$S$};i}(Q^2)_{;i}
    + 2(Q^2)_{;i}(Q^2)_{;i}
  \Bigr]
  \,,
  \\[2mm]
  \tilde{f}^\rmii{$SG$}_3 &=0
  \,,
\end{align}
while the gauge sector is given by
\begin{align}
  \tilde{f}^\rmii{$G$}_3 &=  
      U_{\mu\nu}U_{\nu\sigma}U_{\sigma\mu}
    - \eta_{\mu\mu}Q^2 [\bD_i,[\bD_i,Q^2]]
    - Q^2 [\bD_i,[\bD_i,U_\rmii{$G$}]]
    \nn &
    + \frac{1}{2}U_\rmii{$G$}\bG_{ij}\bG_{ij}
    + \frac{1}{10}U_{\rmii{$G$};iijj}
    + \frac{\eta_{\mu\mu}}{10}(Q^2)_{;iijj} 
    + \frac{\eta_{\mu\mu}}{10}(\bJ_i)^2
    \nn &
    - \frac{\eta_{\mu\mu}}{15}\bG_{ij}\bG_{jk}\bG_{ki}
    + \frac{\eta_{\mu\mu}}{30}[\bD_i,[\bD_j,[\bD_j,\bJ_i]]]
    \nn &
    - \frac{1}{2}[\bD_i, [\bD_i , U_{\mu\nu}U_{\nu\mu}]]
    + \frac{1}{2}\eta_{\mu\mu} [\bD_i, [\bD_i , (Q^2)^2]]
    + [\bD_i, [\bD_i , U_\rmii{$G$} Q^2]]
    \nn &
    + \frac{1}{2}[\bD_i , U_{\mu\nu}][\bD_i , U_{\nu\mu}]]-\frac{1}{2}\eta_{\mu\mu} [\bD_i , Q^2][\bD_i , Q^2]- [\bD_i , Q^2][\bD_i , U_\rmii{$G$}]
    \,,
  \\[2mm]
  \tilde{f}^\rmii{$GS$}_3 &=
    0
  \,.
\end{align}
We note that
$\tilde{f}^\rmii{$SG$}_3$ and
$\tilde{f}^\rmii{$GS$}_3$ are vanishing up to terms beyond dimension six.

\subsection{%
  Direct computation of
  thermal heat kernel coefficients}
\label{app:D-THKC}

In this section,
we present the explicit expressions for
the thermal heat kernel coefficients
$\tilde{C}^{[n,N]}_{X}$
following the strategy
outlined in sec.~\ref{sec:heatKernel:finiteT}
and~\cite{Megias:2003ui}.
Here,
$n$ is the order of the expansion in the proper time, and
$N$ is the number of covariant derivatives, respectively.
The subscript $X \in \{ab,\, 00,\, ij\}$ denotes the field components.

\subsubsection*{Coefficients
  $\widetilde{C}^{[2,N]}_{X}$
  }

\begin{align}
    \widetilde{C}^{[2,0]}_{ab} &= \frac{ \pi^{3/2} \, \g^2}{(\Delta_{12}^2)^2} \Big[
        (Q \bphi_c)(Q \bphi_c) \delta_{ab}
      + (\bD_i \bphi_c)(\bD_i \bphi_c) \delta_{ab}
    \nn &
    \hphantom{{}\frac{ \pi^{3/2} \, \g^2}{(\Delta_{12}^2)^2}\Big[}
      - (Q \bphi_a)(Q \bphi_b)
      - (\bD_i \bphi_a)(\bD_i \bphi_b) \Big]
    \,,\\[1mm]
    \widetilde{C}^{[2,1]}_{ab} &= \pi^{3/2} \, \bD^2 \delta_{ab}  -  \frac{ \pi^{3/2}  \, \g^2}{\Delta_{12}^2}  \Big[(Q \bphi_a)(Q \bphi_b) + (\bD_i \bphi_a)(\bD_i \bphi_b) \Big]
    \,,\\[1mm]
    \widetilde{C}^{[2,2]}_{ab} &= \frac{\pi^{3/2}}{2} \Big[
      (-Q^2 + \bD^2) (-Q^2 + \bD^2) \delta_{ab}
      + U'_{ab} (-Q^2 + \bD^2)
    \nn &
    \hphantom{{}\frac{\pi^{3/2}}{2} \Big[}
      + (-Q^2 + \bD^2) U'_{ab}
      + (U'_{ac} U'_{cb}) \Big] 
    \,,\\[1mm]
    \widetilde{C}^{[2,0]}_{00} &=  \frac{ \pi^{3/2} \, \g^2}{(\Delta_{12}^2)^2} (Q \bphi_a)(Q \bphi_a) 
    \,,\\[1mm]
    \widetilde{C}^{[2,1]}_{00} &=  \pi^{3/2} \, \bD^2 \, \eta_{00} +  \frac{\pi^{3/2} \, \g^2}{\Delta_{12}^2}(Q \bphi_a)(Q \bphi_a)
    \,,\\[1mm]
    \widetilde{C}^{[2,2]}_{00} &= -\frac{\pi^{3/2}}{2} \Big[(-Q^2 + \bD^2) (-Q^2 + \bD^2) \Big] \eta_{00}
    \,,\\[1mm]
    \widetilde{C}^{[2,0]}_{ij} &=  \frac{\pi^{3/2} \, \g^2}{(\Delta_{12}^2)^2} (\bD_i \bphi_a)(\bD_j \bphi_a)
    \,,\\[1mm]
    \widetilde{C}^{[2,1]}_{ij} &= \pi^{3/2} \, \bD^2 \eta_{ij} +  \frac{\pi^{3/2} \, \g^2}{\Delta_{12}^2} (\bD_i \bphi_a)(\bD_j \bphi_a)
    \,,\\[1mm]
    \widetilde{C}^{[2,2]}_{ij} &= \frac{\pi^{3/2} }{2} \Big[
      - ( - Q^2 + \bD^2) (-Q^2 + \bD^2) \eta_{ij}
      + U'_{ij} (-Q^2 + \bD^2)
    \nn &
    \hphantom{{}=\frac{\pi^{3/2}}{2} \Big[}
      + (-Q^2 + \bD^2) U'_{ij}
      + (U'_{ik} U'_{kj})
    \Big] 
    \,.
\end{align}

\subsubsection*{%
  Coefficients
  $\tilde{C}^{[3,N]}_{X}$
  }

\begin{align}
    \widetilde{C}^{[3,0]}_{ab} &= \dfrac{\pi^{3/2} \, \g^2}{(\Delta_{12}^2)^3} \bigg[
        \big( -Q^2 + \bD^2 \big) \Big[(Q \bphi_c) (Q \bphi_c) \delta_{ab} - (Q \bphi_a) (Q \bphi_b) \Big]
    \nn &\hphantom{{}= \dfrac{\pi^{3/2} \, \g^2}{(\Delta_{12}^2)^3} \bigg[}
      + \Big[(Q \bphi_c) (Q \bphi_c) \delta_{ab} - (Q \bphi_a) (Q \bphi_b) \Big] \big(\! -Q^2 + \bD^2 \big)
    \nn &\hphantom{{}= \dfrac{\pi^{3/2} \, \g^2}{(\Delta_{12}^2)^3} \bigg[}
      + U'_{ac} \Big[(Q \bphi_d) (Q \bphi_d) \delta_{cb} - (Q \bphi_c) (Q \bphi_b) \Big]
      + \Big[(Q \bphi_d) (Q \bphi_d) \delta_{ac} - (Q \bphi_a) (Q \bphi_c) \Big] U'_{cb}
    \nn &\hphantom{{}= \dfrac{\pi^{3/2} \, \g^2}{(\Delta_{12}^2)^3} \bigg[}
      + \big( -Q^2 + \bD^2 \big) \Big[(\bD_i \bphi_c) ( \bD_i \bphi_c) \delta_{ab}- (\bD_i \bphi_a) ( \bD_i \bphi_b) \Big]
    \nn &\hphantom{{}= \dfrac{\pi^{3/2} \, \g^2}{(\Delta_{12}^2)^3} \bigg[}
      + \Big[(\bD_i \bphi_c) ( \bD_i \bphi_c) \delta_{ab} - (\bD_i \bphi_a) ( \bD_i \bphi_b) \Big] \big( -Q^2 + \bD^2 \big)
    \nn &\hphantom{{}= \dfrac{\pi^{3/2} \, \g^2}{(\Delta_{12}^2)^3} \bigg[}
      + U'_{ac} \Big[(\bD_i \bphi_d) ( \bD_i \bphi_d) \delta_{cb} - (\bD_i \bphi_c) ( \bD_i \bphi_b) \Big]
    \nn &\hphantom{{}= \dfrac{\pi^{3/2} \, \g^2}{(\Delta_{12}^2)^3} \bigg[}
      + \Big[(\bD_i \bphi_d) ( \bD_i \bphi_d) \delta_{ac} - (\bD_i \bphi_a) ( \bD_i \bphi_c) \Big] U'_{cb}
    \nn &\hphantom{{}= \dfrac{\pi^{3/2} \, \g^2}{(\Delta_{12}^2)^3} \bigg[}
      - 2 \bigl( Q\bphi_{b} \bigr) \bigl( -Q^2 + \bD^2 \bigr) \bigl( Q \bphi_b \bigr) \delta_{ac}
      + 2 \bigl( Q\bphi_{a} \bigr) \bigl( -Q^2 + \bD^2 \bigr) \bigl( Q \bphi_{c} \bigr)
    \nn[1mm] &\hphantom{{}= \dfrac{\pi^{3/2} \, \g^2}{(\Delta_{12}^2)^3} \bigg[}
      - 2 (\bD_i \bphi_b) \big( -Q^2 + \bD^2 \big) (\bD_i \bphi_b) \delta_{ac}
      + 2 (\bD_i \bphi_a) \big( -Q^2 + \bD^2 \big) (\bD_i \bphi_c)
    \nn &\hphantom{{}= \dfrac{\pi^{3/2} \, \g^2}{(\Delta_{12}^2)^3} \bigg[}
      - 2 (\bD_i \bphi_b) U'_{ij} (\bD_j \bphi_b) \delta_{ac}
      + 2 (\bD_i \bphi_a) U'_{ij} (\bD_j \bphi_c)
      \bigg]
    \,,\\[1mm]
    \widetilde{C}^{[3,1]}_{ab} &= - \frac{\pi^{3/2} \,  \g^2}{(\Delta_{12}^2)^2} \bigg[
        \big( -Q^2 + \bD^2 \big) \Big[(Q \bphi_c) (Q \bphi_c) \delta_{ab} - (Q \bphi_a) (Q \bphi_b) \Big]
    \nn &\hphantom{{}= \dfrac{\pi^{3/2} \,  \g^2}{(\Delta_{12}^2)^2} \bigg[}
      + \Big[(Q \bphi_c) (Q \bphi_c) \delta_{ab} - (Q \bphi_a) (Q \bphi_b) \Big] \big(\! -Q^2 + \bD^2 \big)
    \nn &\hphantom{{}= \dfrac{\pi^{3/2} \,  \g^2}{(\Delta_{12}^2)^2} \bigg[}
      + U'_{ac} \Big[(Q \bphi_d) (Q \bphi_d) \delta_{cb} - (Q \bphi_c) (Q \bphi_b) \Big]
      + \Big[(Q \bphi_d) (Q \bphi_d) \delta_{ac} - (Q \bphi_a) (Q \bphi_c) \Big] U'_{cb}
    \nn &\hphantom{{}= \dfrac{\pi^{3/2} \,  \g^2}{(\Delta_{12}^2)^2} \bigg[}
      + \big( -Q^2 + \bD^2 \big) \Big[(\bD_i \bphi_c) ( \bD_i \bphi_c) \delta_{ab}- (\bD_i \bphi_a) ( \bD_i \bphi_b) \Big]
    \nn &\hphantom{{}= \dfrac{\pi^{3/2} \,  \g^2}{(\Delta_{12}^2)^2} \bigg[}
      + \Big[(\bD_i \bphi_c) ( \bD_i \bphi_c) \delta_{ab} - (\bD_i \bphi_a) ( \bD_i \bphi_b) \Big] \big( -Q^2 + \bD^2 \big)
    \nn &\hphantom{{}= \dfrac{\pi^{3/2} \,  \g^2}{(\Delta_{12}^2)^2} \bigg[}
      + U'_{ac} \Big[(\bD_i \bphi_d) ( \bD_i \bphi_d) \delta_{cb} - (\bD_i \bphi_c) ( \bD_i \bphi_b) \Big]
    \nn &\hphantom{{}= \dfrac{\pi^{3/2} \,  \g^2}{(\Delta_{12}^2)^2} \bigg[}
      + \Big[(\bD_i \bphi_d) ( \bD_i \bphi_d) \delta_{ac}- (\bD_i \bphi_a) ( \bD_i \bphi_c) \Big] U'_{cb}
    \nn &\hphantom{{}= \dfrac{\pi^{3/2} \,  \g^2}{(\Delta_{12}^2)^2} \bigg[}
      - \bigl( Q\bphi_{b} \bigr) \bigl( -Q^2 + \bD^2 \bigr) \bigl( Q \bphi_b \bigr) \delta_{ac}
      + \bigl( Q\bphi_{a} \bigr) \bigl( -Q^2 + \bD^2 \bigr) \bigl( Q \bphi_{c} \bigr)
    \nn[1mm] &\hphantom{{}= \dfrac{\pi^{3/2} \,  \g^2}{(\Delta_{12}^2)^2} \bigg[}
      - (\bD_i \bphi_b) \big( -Q^2 + \bD^2 \big) (\bD_i \bphi_b) \delta_{ac}
      + (\bD_i \bphi_a) \big( -Q^2 + \bD^2 \big) (\bD_i \bphi_c)
    \nn &\hphantom{{}= \dfrac{\pi^{3/2} \,  \g^2}{(\Delta_{12}^2)^2} \bigg[}
      - (\bD_i \bphi_b) U'_{ij} (\bD_j \bphi_b) \delta_{ac}
      + (\bD_i \bphi_a) U'_{ij} (\bD_j \bphi_c)
    \bigg]
    \,,\\[1mm]
    \widetilde{\mathbb{C}}^{[3,1]}_{ab} &= \frac{\pi^{3/2} \, \g^2}{(\Delta_{12}^2)^2} \bigg[
        \left( Q\bphi_{b} \right) \left( -Q^2 + \bD^2 \right) \left( Q \bphi_b \right) \delta_{ac}
      - \left( Q\bphi_{a} \right) \left( -Q^2 + \bD^2 \right) \left( Q \bphi_{c} \right)
    \nn[1mm] &\hphantom{{}= \dfrac{\pi^{3/2} \, \g^2}{(\Delta_{12}^2)^2} \bigg[}
      + (\bD_i \bphi_b) \big( -Q^2 + \bD^2 \big) (\bD_i \bphi_b) \delta_{ac}
      - (\bD_i \bphi_a) \big( -Q^2 + \bD^2 \big) (\bD_i \bphi_c)
    \nn[1mm] &\hphantom{{}= \dfrac{\pi^{3/2} \, \g^2}{(\Delta_{12}^2)^2} \bigg[}
      + (\bD_i \bphi_b) U'_{ij} (\bD_j \bphi_b) \delta_{ac}
      - (\bD_i \bphi_a) U'_{ij} (\bD_j \bphi_c)
    \bigg]
    \,,\\[1mm]
    \widetilde{C}^{[3,2]}_{ab} &= \frac{\pi^{3/2}}{3} \bigg[
        \bD^2 \big[- Q^2 + \bD^2 \big] \, \delta_{ab}
      - \bD_i \big[ -Q^2 + \bD^2 \big] \bD_i \, \delta_{ab}
      + \big[ -Q^2 + \bD^2 \big] \bD^2 \, \delta_{ab}
      \nn &\hphantom{{}=\frac{\pi^{3/2}}{3} \bigg[}
      + \bD^2 \, U'_{ab}
      - \bD_i U'_{ab} \bD_i
      + U'_{ab} \bD^2 \bigg]
    \nn &
      - \dfrac{\pi^{3/2} \, \g^2}{2 \, \Delta_{12}^2 } \bigg[
          \big( -Q^2 + \bD^2 \big) \Big[(Q \bphi_c) (Q \bphi_c) \delta_{ab} - (Q \bphi_a) (Q \bphi_b) \Big]
    \nn &\hphantom{{}= \dfrac{\pi^{3/2} \, \g^2}{2 \, \Delta_{12}^2 } \bigg[}
        + \Big[(Q \bphi_c) (Q \bphi_c) \delta_{ab} - (Q \bphi_a) (Q \bphi_b) \Big] \big(\! -Q^2 + \bD^2 \big)
    \nn &\hphantom{{}= \dfrac{\pi^{3/2} \, \g^2}{2 \, \Delta_{12}^2 } \bigg[}
      + U'_{ac} \Big[(Q \bphi_d) (Q \bphi_d) \delta_{cb} - (Q \bphi_c) (Q \bphi_b) \Big]
      + \Big[(Q \bphi_d) (Q \bphi_d) \delta_{ac} - (Q \bphi_a) (Q \bphi_c) \Big] U'_{cb}
    \nn &\hphantom{{}= \dfrac{\pi^{3/2} \, \g^2}{2 \, \Delta_{12}^2 } \bigg[}
      + \big( -Q^2 + \bD^2 \big) \Big[(\bD_i \bphi_c) ( \bD_i \bphi_c) \delta_{ab}
      - (\bD_i \bphi_a) ( \bD_i \bphi_b) \Big]
    \nn &\hphantom{{}= \dfrac{\pi^{3/2} \, \g^2}{2 \, \Delta_{12}^2 } \bigg[}
      + \Big[(\bD_i \bphi_c) ( \bD_i \bphi_c) \delta_{ab} - (\bD_i \bphi_a) ( \bD_i \bphi_b) \Big] \big( -Q^2 + \bD^2 \big)
    \nn &\hphantom{{}= \dfrac{\pi^{3/2} \, \g^2}{2 \, \Delta_{12}^2 } \bigg[}
      + U'_{ac} \Big[(\bD_i \bphi_d) ( \bD_i \bphi_d) \delta_{cb} - (\bD_i \bphi_c) ( \bD_i \bphi_b) \Big]
    \nn &\hphantom{{}= \dfrac{\pi^{3/2} \, \g^2}{2 \, \Delta_{12}^2 } \bigg[}
      + \Big[(\bD_i \bphi_d) ( \bD_i \bphi_d) \delta_{ac} - (\bD_i \bphi_a) ( \bD_i \bphi_c) \Big] U'_{cb}
    \bigg]
    \,,\\[1mm]
    \widetilde{C}^{[3,3]}_{ab} &= \frac{\pi^{3/2}}{3!} \bigg[
        \big(-Q^2 + \bD^2 \big)^3 \delta_{ab}
      + (U')^3_{ab}
      + \big(-Q^2 + \bD^2 \big)^2 U'_{ab}
      \nn &\hphantom{{}= \frac{\pi^{3/2}}{3!} \bigg[}
      + \big(-Q^2 + \bD^2 \big) U'_{ab} \big(-Q^2 + \bD^2 \big)
      + U'_{ab} \big(-Q^2 + \bD^2 \big)^2
    \\ &\hphantom{{}= \frac{\pi^{3/2}}{3!} \bigg[}
      + (U')^2_{ab} \big(-Q^2 + \bD^2 \big)
      + (U')_{ac} \big(-Q^2 + \bD^2 \big) (U')_{cb}
      + \big(-Q^2 + \bD^2 \big) (U')^2_{ab}
    \bigg]
    \,,\nn[1mm]
    \widetilde{C}^{[3,0]}_{00} &= \dfrac{\pi^{3/2} \, \g^2}{(\Delta_{12}^2)^3}
    \biggl[
        2 (Q \bphi_a)(\! -Q^2 + \bD^2)(Q \bphi_a)
      + 2 (Q \bphi_a)U'_{bb}(Q \bphi_a)
      - 2 (Q \bphi_a)U'_{ac}(Q \bphi_c)
    \nn &\hphantom{{}= \dfrac{\pi^{3/2} \, \g^2}{(\Delta_{12}^2)^3} \Big[}
      - (Q \bphi_a)(Q \bphi_a) \big(\! - Q^2 + \bD^2 \big)
      - \big(\! -Q^2 + \bD^2 \big) (Q \bphi_a)(Q \bphi_a)
    \biggr]
    \,,\\[1mm]
    \widetilde{C}^{[3,1]}_{00} &= \dfrac{\pi^{3/2} \, \g^2}{(\Delta_{12}^2)^2} \biggl[
        (Q \bphi_a)(\! -Q^2 + \bD^2)(Q \bphi_a)
      + (Q \bphi_a)U'_{bb}(Q \bphi_a)
      - (Q \bphi_a)U'_{ac}(Q \bphi_c)
    \nn &\hphantom{{}= \dfrac{\pi^{3/2} \, \g^2}{(\Delta_{12}^2)^2} \biggl[}
      - (Q \bphi_a)(Q \bphi_a) \big(\! - Q^2 + \bD^2 \big)
      - \big(\! -Q^2 + \bD^2 \big) (Q \bphi_a)(Q \bphi_a)
    \biggr]
    \,,\\[1mm]
    \widetilde{\mathbb{C}}^{[3,1]}_{00} &= \dfrac{\pi^{3/2} \, \g^2}{(\Delta_{12}^2)^2} \biggl[
        (Q \bphi_a)(\! -Q^2 + \bD^2)(Q \bphi_a)
      + (Q \bphi_a)U'_{bb}(Q \bphi_a)
      - (Q \bphi_a)U'_{ac}(Q \bphi_c)
    \biggr]
    \,,\\[1mm],
    \widetilde{C}^{[3,2]}_{00} &= \frac{\pi^{3/2}}{3} \bigg[
          \bD^2 \big(-Q^2 + \bD^2 \big)
        - \bD_i \big(-Q^2 + \bD^2 \big) \bD_i
        + \big(-Q^2 + \bD^2 \big) \bD^2
      \bigg]
    \nn &
      + \dfrac{\pi^{3/2} \, \g^2}{2 \, \Delta_{12}^2} \bigg[
          (Q \bphi_a)(Q \bphi_a) \big(\! - Q^2 + \bD^2 \big)
        + \big(\! -Q^2 + \bD^2 \big) (Q \bphi_a)(Q \bphi_a)
      \bigg]
    \,,\\[1mm]
    \widetilde{C}^{[3,3]}_{00} &= \frac{\pi^{3/2}}{3!} \big(-Q^2 + \bD^2 \big)^3
    \,,\\[1mm]
    \widetilde{C}^{[3,0]}_{ij} &= \dfrac{\pi^{3/2} \, \g^2}{(\Delta_{12}^2)^3} \biggl[
        2 (\bD_i \bphi_a) \big( - Q^2 + \bD^2 \big) (\bD_j \bphi_a)
      + 2 (\bD_i \bphi_a) U'_{bb} (\bD_j \bphi_a)
      - 2 (\bD_i \bphi_a) U'_{ab} (\bD_j \bphi_b)
    \nn &\hphantom{{}= \dfrac{\pi^{3/2} \, \g^2}{(\Delta_{12}^2)^3} \bigg[}
      - (\bD_i \bphi_a) (\bD_j \bphi_a) \big( - Q^2 + \bD^2 \big)
      - (\bD_i \bphi_a) (\bD_k \bphi_a) U'_{kj}
    \nn[1mm] &\hphantom{{}= \dfrac{\pi^{3/2} \, \g^2}{(\Delta_{12}^2)^3} \Big[}
      - \big( - Q^2 + \bD^2 \big) (\bD_i \bphi_a) (\bD_j \bphi_a)
      - U'_{ik} (\bD_k \bphi_a) (\bD_j \bphi_a)
    \biggr]
    \,,\\[1mm]
    \widetilde{C}^{[3,1]}_{ij} &= \dfrac{\pi^{3/2} \, \g^2}{(\Delta_{12}^2)^2} \biggl[
        (\bD_i \bphi_a) \big( - Q^2 + \bD^2 \big) (\bD_j \bphi_a)
      + (\bD_i \bphi_a) U'_{bb} (\bD_j \bphi_a)
      - (\bD_i \bphi_a) U'_{ab} (\bD_j \bphi_b)
    \nn &\hphantom{{}= \dfrac{\pi^{3/2} \, \g^2}{(\Delta_{12}^2)^2} \biggl[}
      - (\bD_i \bphi_a) (\bD_j \bphi_a) \big( - Q^2 + \bD^2 \big)
      - (\bD_i \bphi_a) (\bD_k \bphi_a) U'_{kj}
    \nn[1mm] &\hphantom{{}= \dfrac{\pi^{3/2} \, \g^2}{(\Delta_{12}^2)^2} \biggl[}
      - \big( - Q^2 + \bD^2 \big) (\bD_i \bphi_a) (\bD_j \bphi_a)
      - U'_{ik} (\bD_k \bphi_a) (\bD_j \bphi_a)
      \biggr]
    \,,\\[1mm]
    \widetilde{ \mathbb{C}}^{[3,1]}_{ij} &= \dfrac{\pi^{3/2} \, \g^2}{(\Delta_{12}^2)^2} \biggl[
        (\bD_i \bphi_a) \big( - Q^2 + \bD^2 \big) (\bD_j \bphi_a)
      + (\bD_i \bphi_a) U'_{bb} (\bD_j \bphi_a)
      - (\bD_i \bphi_a) U'_{ab} (\bD_j \bphi_b)
    \biggr]
    \,,\\[1mm]
    \widetilde{C}^{[3,2]}_{ij} &= -\frac{\pi^{3/2}}{3} \biggl[
      - \bD_k \big(-Q^2 + \bD^2 \big) \bD_k \eta_{ij}
      + \big(-Q^2 + \bD^2 \big) \bD^2 \eta_{ij}
      \nn &\hphantom{{}= -\frac{\pi^{3/2}}{3} \biggl[}
      + \bD^2 U'_{ij}
      - \bD_k U'_{ij} \bD_k
      + U'_{ij} \bD^2
    \biggr]
    \nn &
      + \frac{\pi^{3/2} \, \g^2}{ 2 \, \Delta_{12}^2 } \biggl[
          (\bD_i \bphi_a) (\bD_j \bphi_a) \big( - Q^2 + \bD^2 \big)
        + (\bD_i \bphi_a) (\bD_k \bphi_a) U'_{kj}
      \nn &\hphantom{{}= \frac{\pi^{3/2} \, \g^2}{ 2 \, \Delta_{12}^2 } \biggl[}
        + \big( - Q^2 + \bD^2 \big) (\bD_i \bphi_a) (\bD_j \bphi_a)
        + U'_{ik} (\bD_k \bphi_a) (\bD_j \bphi_a)
      \biggr]
    \,,\\[1mm]
    \widetilde{C}^{[3,3]}_{ij} &= \frac{\pi^{3/2}}{3!} \biggl[
      -\big(-Q^2 + \bD^2 \big)^3 \eta_{ij}
      + (U')^3_{ij}
      + \big(-Q^2 + \bD^2 \big)^2 U'_{ij}
      \nn &\hphantom{{}= \frac{\pi^{3/2}}{3!} \biggl[}
      + \big(-Q^2 + \bD^2 \big) U'_{ij} \big(-Q^2 + \bD^2 \big)
      + U'_{ij} \big(-Q^2 + \bD^2 \big)^2
    \\ &\hphantom{{}= \frac{\pi^{3/2}}{3!} \bigg[}
      + (U')^2_{ij} \big(-Q^2 + \bD^2 \big)
      + (U')_{ik} \big(-Q^2 + \bD^2 \big) (U')_{kj}
      + \big(-Q^2 + \bD^2 \big) (U')^2_{ij}
    \biggr]
    \,.\nonumber
\end{align}

\subsubsection*{%
  Coefficients
  $\tilde{C}^{[4,N]}_{X}$
  }
\begin{align}
    \widetilde{C}^{[4,3]}_{ab} &= \frac{1}{3!} \bigg[
      \bD^2 \Big((-Q^2 + \bD^2)\delta_{ad} + U'_{ad}\Big)
      \Big((-Q^2 + \bD^2)\delta_{db} + U'_{db}\Big)
    \nn &\hphantom{{}= \frac{1}{3!} \bigg[}
      + \bD_i \Big((-Q^2 + \bD^2)\delta_{ac} + U'_{ac}\Big) \bD_i
      \Big((-Q^2 + \bD^2)\delta_{cb} + U'_{cb}\Big)
    \nn &\hphantom{{}= \frac{1}{3!} \bigg[}
      + \bD_i \Big((-Q^2 + \bD^2)\delta_{ac} + U'_{ac}\Big)
      \Big((-Q^2 + \bD^2)\delta_{cb} + U'_{cb}\Big) \bD_i
    \nn &\hphantom{{}= \frac{1}{3!} \bigg[}
      + \Big((-Q^2 + \bD^2)\delta_{ac} + U'_{ac}\Big)
      \bD_i \Big((-Q^2 + \bD^2)\delta_{cb} + U'_{cb}\Big) \bD_i
    \nn &\hphantom{{}= \frac{1}{3!} \bigg[}
      + \Big((-Q^2 + \bD^2)\delta_{ac} + U'_{ac}\Big)
      \Big((-Q^2 + \bD^2)\delta_{cb} + U'_{cb}\Big) \bD^2
    \bigg]
    \,,\\[1mm]
    \widetilde{C}^{[4,3]}_{00} &= \frac{1}{3!} \bigg[
      \bD^2 \big(-Q^2 + \bD^2\big) \big(-Q^2 + \bD^2\big)
      + \bD_i \big(-Q^2 + \bD^2\big) \bD_i \big(-Q^2 + \bD^2\big)
    \nn &\hphantom{{}= \frac{1}{3!} \bigg[}
      + \bD_i \big(-Q^2 + \bD^2\big) \big(-Q^2 + \bD^2\big) \bD_i
      + \big(-Q^2 + \bD^2\big) \bD^2 \big(-Q^2 + \bD^2\big)
    \nn &\hphantom{{}= \frac{1}{3!} \bigg[}
      + \big(-Q^2 + \bD^2\big) \big(-Q^2 + \bD^2\big) \bD^2
    \bigg]
    \,,\\[1mm]
    \widetilde{C}^{[4,3]}_{ij} &= \frac{1}{3!} \bigg[
      \bD^2 \Big((-Q^2 + \bD^2)\eta_{il} + U'_{il}\Big)
      \Big((-Q^2 + \bD^2)\eta_{lj} + U'_{lj}\Big)
    \nn &\hphantom{{}= \frac{1}{3!} \bigg[}
      + \bD_k \Big((-Q^2 + \bD^2)\eta_{il} + U'_{il}\Big) \bD_k
      \Big((-Q^2 + \bD^2)\eta_{lj} + U'_{lj}\Big)
    \nn &\hphantom{{}= \frac{1}{3!} \bigg[}
      + \bD_k \Big((-Q^2 + \bD^2)\eta_{il} + U'_{il}\Big)
      \Big((-Q^2 + \bD^2)\eta_{lj} + U'_{lj}\Big) \bD_k
    \nn &\hphantom{{}= \frac{1}{3!} \bigg[}
      + \Big((-Q^2 + \bD^2)\eta_{il} + U'_{il}\Big) \bD_k
      \Big((-Q^2 + \bD^2)\eta_{lj} + U'_{lj}\Big) \bD_k
    \nn &\hphantom{{}= \frac{1}{3!} \bigg[}
      + \Big((-Q^2 + \bD^2)\eta_{il} + U'_{il}\Big)
      \Big((-Q^2 + \bD^2)\eta_{lj} + U'_{lj}\Big) \bD^2
    \bigg]
    \,.
\end{align}

\subsubsection*{%
  Coefficients
  $\widetilde{C}^{[5,N]}_{X}$
  }

\begin{align}
    \widetilde{C}^{[5,3]}_{ab} &= \frac{\pi^{3/2}}{30} \bigg[
      \big(\bD_i\bD_i\bD_j\bD_j + \bD_i\bD_j\bD_j\bD_i + \bD_i\bD_j\bD_i\bD_j \big)
      \,\big((-Q^2 + \bD^2)\delta_{ab} + U'_{ab}\big)
    \nn &\hphantom{{}= \frac{\pi^{3/2}}{30} \bigg[}
      + \big(\bD_i\bD_i\bD_j + \bD_i\bD_j\bD_i \big)
        \,\big((-Q^2 + \bD^2)\delta_{ab} + U'_{ab}\big)\,\bD_j
    \nn[1mm] &\hphantom{{}= \frac{\pi^{3/2}}{30} \bigg[}
      + \bD_i\bD_j\bD_j
        \big((-Q^2 + \bD^2)\delta_{ab} + U'_{ab}\big)\, \bD_i
      + \bD^2 \,\big((-Q^2 + \bD^2)\delta_{ab} + U'_{ab}\big)\, \bD^2
    \nn[1mm] &\hphantom{{}= \frac{\pi^{3/2}}{30} \bigg[}
      + \bD_j\bD_i \,\big((-Q^2 + \bD^2)\delta_{ab}\bD_i\bD_j + U'_{ab}\big)
      + \bD_i\bD_j \,\big((-Q^2 + \bD^2)\delta_{ab} + U'_{ab}\big)\, \bD_j\bD_i
    \nn[1mm] &\hphantom{{}= \frac{\pi^{3/2}}{30} \bigg[}
      + \bD_i \,\big((-Q^2 + \bD^2)\delta_{ab} + U'_{ab}\big)
        \,\big(\bD_i\bD_j\bD_j + \bD_j\bD_j\bD_i + \bD_j\bD_i\bD_j \big)
    \nn &\hphantom{{}= \frac{\pi^{3/2}}{30} \bigg[}
      + \big((-Q^2 + \bD^2)\delta_{ab} + U'_{ab}\big)
        \,\big(\bD_i\bD_i\bD_j\bD_j + \bD_i\bD_j\bD_j\bD_i
        + \bD_i\bD_j\bD_i\bD_j \big)
    \bigg]
    \,,\\[1mm]
    \widetilde{C}^{[5,3]}_{00} &= \frac{\pi^{3/2}}{10} \bigg[
      \bD^4 \,\big(-Q^2 + \bD^2\big)
      + \bD^2 \bD_i \,\big(-Q^2 + \bD^2\big)\,\bD_i
      + \bD^2 \,\big(-Q^2 + \bD^2\big)\,\bD^2
    \nn &\hphantom{{}= \frac{\pi^{3/2}}{10} \bigg[}
      + \bD_i \,\big(-Q^2 + \bD^2\big)\,\bD_i \bD^2
      + \big(-Q^2 + \bD^2\big)\,\bD^4
    \bigg]
    \,,\\[1mm]
    \widetilde{C}^{[5,3]}_{ij} &= \frac{\pi^{3/2}}{10} \bigg[
      \bD^4 \,\big((-Q^2 + \bD^2)\eta_{ij} + U'_{ij}\big)
      + \bD^2 \bD_i \,\big((-Q^2 + \bD^2)\eta_{ij} + U'_{ij}\big)\,\bD_i
    \nn &\hphantom{{}= \frac{\pi^{3/2}}{10} \bigg[}
      + \bD^2 \,\big((-Q^2 + \bD^2)\eta_{ij} + U'_{ij}\big)\,\bD^2
      + \bD_i \,\big((-Q^2 + \bD^2)\eta_{ij} + U'_{ij}\big)\,\bD_i \bD^2
    \nn &\hphantom{{}= \frac{\pi^{3/2}}{10} \bigg[}
      + \big((-Q^2 + \bD^2)\eta_{ij} + U'_{ij}\big)\,\bD^4
    \bigg]
    \,.
\end{align}

\subsubsection*{%
  Coefficients
  $\widetilde{C}^{[6,N]}_{X}$
  }

\begin{align}
    \widetilde{C}^{[6,3]}_{ab} &= \frac{\pi^{3/2}}{90}
      \big(
        \bD_i\bD_i\bD_j\bD_j\bD_k\bD_k
      + \text{14 permutations}
    \big) \delta_{ab} 
    \,, \\[1mm]
    \widetilde{C}^{[6,3]}_{00} &= \frac{\pi^{3/2}}{90} \big(
        \bD_i\bD_i\bD_j\bD_j\bD_k\bD_k
      + \text{14 permutations}
      \big) 
    \,, \\[1mm]
    \widetilde{C}^{[6,3]}_{ij} &= - \frac{\pi^{3/2}}{90} \big(
        \bD_k\bD_k\bD_l\bD_l\bD_m\bD_m
      + \text{14 permutations}
      \big) \eta_{ij} 
    \,.
\end{align}

{\small
\bibliographystyle{utphys}
\bibliography{ref}
}
\end{document}